\documentstyle[epsfig,12pt]{article}

\input{tcilatex}

\begin{document}

\begin{titlepage}
\bigskip \begin{flushright}
WATPHYS-TH01/02
\end{flushright}



\vspace{1cm}

\begin{center}
{\Large \bf{Quasilocal Thermodynamics of Kerr and Kerr-anti-de Sitter
Spacetimes }}
\vspace{1.5cm}
{\Large \bf{and the AdS/CFT Correspondence}} 
\end{center}
\vspace{2cm}
\begin{center}
M.H. Dehghani\footnote{%
Electronic address: hossein@avatar.uwaterloo.ca; On leave from Physics
Dept., College of Sciences, Shiraz University, Shiraz, Iran} and R. B. Mann{%
\footnote{%
EMail: mann@avatar.uwaterloo.ca}} \\
Department of Physics, University of Waterloo, \\
Waterloo, Ontario N2L 3G1, CANADA\\
\vspace{2cm}
\vspace{2cm}
\today\\
\end{center}

\begin{abstract}
We consider the quasilocal thermodynamics of rotating black holes in
asymptotically flat and asymptotically anti de Sitter (AdS) spacetimes. \ Using
the minimal number of intrinsic boundary counterterms inspired by the
AdS/conformal field theory correspondence, we find that we are able to carry out 
an analysis of
the thermodynamics of these black holes for virtually all possible values of
the rotation parameter and cosmological constant that leave the quasilocal
boundary well defined, going well beyond what is possible with background
subtraction methods. Specifically, we compute the quasilocal energy $E$ and
angular momentum $J$ for arbitrary values of the rotation, mass and
cosmological constant parameters for the (3+1)- dimensional Kerr, Kerr-AdS
black holes and (2+1)- dimensional Banados Teittelboim and Zanelli (BTZ) black hole.
 We perform a quasilocal stability analysis and find phase behavior that is
commensurate with previous analyses carried out at infinity.
\end{abstract}
\end{titlepage}\onecolumn


\section{Introduction}

Asymptotically anti-de Sitter (AdS) spacetimes continue to attract a great deal of
attention, primarily because of the recently conjectured AdS conformal field
theory (CFT) correspondence principle . This principle posits a relationship
between supergravity or string theory in bulk AdS spacetimes and conformal
field theories on the boundary. It offers the tantalizing possibility that a
full quantum theory of gravity could be described by a well understood
CFT-Yang-Mills theory (albeit with the asymptotic AdS criterion).
Observables in the quantum gravity theory could be computed using quantum
field-theoretic methods, and vice versa. \ Although there is no proof of the
conjecture, a dictionary is emerging that translates between different
quantities in the bulk gravity theory and its counterparts on the boundary,
including the partition functions and correlation functions of each. A set
of quantities of particular interest are those associated with gravitational
thermodynamics. \ It has long been known \ that a physical entropy $S$ and
temperature $\beta ^{-1}$ can be ascribed to a given black hole
configuration, where these quantities are respectively proportional to the
area and surface gravity of the event horizon(s) \cite{beck,hawk1}. Other
black hole properties, such as energy, angular momentum and conserved charge
can also be given a thermodynamic interpretation \cite{gibh}. More recently
it has been shown that entropy can be associated with a broader and
qualitatively different object called a Misner string, which is the
gravitational analogue of a Dirac string \cite
{Hawkhunt,nutent,EJM,HPH,nutkerr,Fatibene}. \ 

In general the problem of defining and calculating gravitational
thermodynamic quantities remains a lively subject of interest. Because they
typically diverge for both asymptotically flat and asymptotically AdS
spacetimes, a common approach toward evaluating them has been to carry out
all computations relative to some other spacetime that is regarded as the
ground state for the class of spacetimes of interest. \ This is done by
taking the original action $I_{G}=I_{{\cal M}}+I_{\partial {\cal M}}$\ for
gravity coupled to matter fields and subtracting from it a reference action $%
I_{0}$, which is a functional of the induced metric $\gamma $ on the
boundary $\partial {\cal M}$. \ \ Conserved and/or thermodynamic quantities
are then computed relative to this boundary, which can then be taken to
(spatial) infinity if desired. \ 

This approach has been widely successful in providing a description of
gravitational thermodynamics in regions of both finite and infinite spatial
extent \cite{BY,BCM}. Unfortunately it suffers from several drawbacks. The
choice of reference spacetime is not always unique \cite{CCM}, nor is it
always possible to embed a boundary with a given induced metric into the
reference background. Indeed, for Kerr spacetimes, this latter problem forms
a serious obstruction towards calculating the subtraction energy, and
calculations have only been performed in the slow-rotating regime \cite
{Martinez}.

An extension of this approach, which addresses these difficulties was
recently developed based on the conjectured AdS/CFT correspondence \cite
{hensken,hyun,balakraus,sergey}. Since quantum field theories in general
contain counterterms, it is natural from the AdS/CFT viewpoint to append a
boundary term $I_{ct}$\ to the action that depends on the intrinsic geometry
of the (timelike) boundary at large spatial distances. This requirement,
along with general covariance, implies that these terms be functionals of
curvature invariants of the induced metric and have no dependence on the
extrinsic curvature of the boundary. An algorithmic procedure \cite{kls}
exists for constructing $I_{ct}$\ \ for asymptotically AdS spacetimes, and
so its determination is unique. The addition of $I_{ct}$\ will not affect
the bulk equations of motion, thereby eliminating the need to embed the
given geometry in a reference spacetime. Hence thermodynamic and conserved
quantities can now be calculated intrinsically for any given spacetime. The
efficacy of this approach has been demonstrated in a broad range of
examples, all of them in the spatially infinite limit, where the AdS/CFT
correspondence applies \cite{nutent,EJM,nutkerr,Awad,vish}. \ However this
limit is an idealization, and can never be satisfied in reality.

The purpose of this paper is to investigate the effects of including $I_{ct}$%
\ for quasilocal gravitational thermodynamics for situations in which the
region enclosed by \ $\partial {\cal M}$ is spatially finite. There are
several reasons for considering this. Although the AdS/CFT correspondence is
conjectured to be valid only at infinity, \ from a gravitational viewpoint
the inclusion of additional boundary functionals is not uniquely dependent
upon this correspondence and could in fact be carried out even if the
conjecture is ultimately found to be invalid. It is, therefore of interest
to explore the implications of $I_{ct}$\ outside of its original conceptual
framework. \ Furthermore, the inclusion of $I_{ct}$\ eliminates the
embedding problem from consideration whether or not the spatially infinite
limit is taken, and it is of interest to see what its impact is on
quasilocal thermodynamics. Since $I_{ct}$ is a functional of the induced
metric, it will in general depend upon parameters in the solution related to
conserved quantities at infinity, and could conceivably alter the results
obtained for quasilocal thermodynamics relative to a fixed background. \ Of
course $I_{ct}$\ itself is no longer unique, since one could add any
functional of curvature invariants that vanished at infinity in a given
dimension. However the expression for $I_{ct}$\ obtained by the algorithm
given in Ref. \cite{kls} is the minimal one required to eliminate
divergences, and therefore merits study.

We specifically consider in this paper an investigation of the thermodynamic
properties of the class of Kerr and Kerr AdS black holes in the context of
spatially finite boundary conditions, with the boundary action supplemented
by $I_{ct}$. With their lower degree of symmetry relative to spherically
symmetric black holes, these spacetimes allow for a more detailed study of
the consequences of including $I_{ct}$ for spatially finite boundaries. \
Although the inclusion of $I_{ct}$ was originally proposed for
asymptotically AdS spacetimes, it has been demonstrated that it can be
extended to asymptotically flat spacetimes \cite{nutent,nutkerr,lau} under
certain topological constraints \cite{kls}. \ We consider the implications
of this extension for Kerr black holes in spatially finite regions.

The outline of our paper is as follows. We review the basic formalism in
Sec. \ref{Genfor}. In the next two sections we compute the quasilocal
energy, mass, and angular momentum for the Kerr and Kerr-anti-de Sitter
black holes respectively. These quantities must be computed numerically for
general values of the parameters, and we present our results graphically. We
perform a simple stability analysis for each case, and compare this to
previous analyses carried out at infinity \cite{Davies,Cald}. We finish our
paper with some concluding remarks.

\section{General Formalism\label{Genfor}}

\bigskip The action is the sum of three kinds of terms. \ The first two are
a volume (or bulk) term 
\begin{equation}
I_{v}=-\frac{1}{16\pi }\int_{{\cal M}}d^{4}x\sqrt{-g}[{\cal R}-2\Lambda +%
{\cal L}(\Phi )]  \label{e3}
\end{equation}
and a boundary term \ 
\begin{equation}
I_{b}=-\frac{1}{8\pi }\int_{\partial {\cal M}}d^{3}x\sqrt{-\gamma }\Theta
(\gamma )  \label{e4}
\end{equation}
chosen so that the variational principle is well defined. The Euclidean
manifold ${\cal M}$ has the metric $g_{\mu \nu }$, covariant derivative $%
\nabla _{\mu }$, and time coordinate $\tau $ which foliates ${\cal M}$ into
non-singular hypersurfaces $\Sigma _{\tau }$ with unit normal $u_{\mu }$
over a real line interval $\Upsilon $. $\Theta $ is the trace of the
extrinsic curvature $\Theta ^{\mu \nu }$ of any boundary(ies) $\partial 
{\cal M}$ of the manifold ${\cal M}$, with the induced metric(s) $\gamma
_{ij}$; the manifold can have internal boundary components as well as a
boundary at infinity, although only the latter will be needed in what
follows. Here ${\cal L}(\Phi )$ is the matter Lagrangian and $\Lambda $ the
cosmological constant, which, in what follows, will be taken to be either
zero or negative.

In general $I_{v}$ and $I_{b}$ are both divergent when evaluated on
solutions, as is the Hamiltonian, and other associated thermodynamic
quantities \cite{BY,BCM}. \ Rather than eliminating these divergences by
incorporating a reference term in the spacetime \cite{BCM,ivan}, a new term $%
\ I_{ct}$ is added to the action which is a functional only of boundary
curvature invariants. Although there are a very large number of possible
invariants one could add in a given dimension, only a finite number of them
can eliminate the divergences at spatial infinity, which arise from $I_{v}$
and $I_{b}$. For an asymptotically AdS spacetime, these can be determined by
an algorithmic procedure \cite{kls}. \ Quantities such as energy, entropy,
etc., can then be understood as intrinsically defined for a given spacetime,
as opposed to being defined relative to some abstract (and non-unique)
background, although this latter option is still available. In four
dimensions the algorithm yields 
\begin{equation}
I_{ct(AdS)}=\frac{2}{l}\frac{1}{8\pi }\int_{\partial {\cal M}_{\infty
}}d^{3}x\sqrt{-\gamma }\left( 1+\frac{l^{2}}{4}{\cal R}(\gamma )\right) ,
\label{e5}
\end{equation}
where $l^{2}=3/|\Lambda |$ and the coefficients have been chosen \cite
{nutent,nutkerr,kls} so that the total action $I=I_{v}+I_{b}+I_{ct(Ads)}$ is
finite. We shall study the implications of including the terms in (\ref{e5})
for Kerr-AdS black holes in spatially finite regions. Although other
counterterms (of higher mass dimension) may be added to $I_{ct}$, they will
make no contribution to the evaluation of the action or Hamiltonian due to
the rate at which they decrease toward spatial infinity, and we shall not
consider them in our analysis here.

\bigskip A generalization of the prescription (\ref{e5}) to spacetimes that
are not asymptotically AdS is \cite{nutent,nutkerr} 
\begin{equation}
I_{ct}=\frac{2}{l}\frac{1}{8\pi }\int_{\partial {\cal M}_{\infty }}d^{3}x%
\sqrt{-\gamma }\sqrt{1+\frac{l^{2}}{2}{\cal R}(\gamma )},  \label{e7}
\end{equation}
which is equivalent to Eq. (\ref{e5}) for small $l$ (as well as fixed $l$
and large mean boundary radius), and which (formally) has a well defined
limit for vanishing cosmological constant ($l\rightarrow \infty $). A
similar formula, in which ${\cal R}(\gamma )$ is replaced with the Ricci
scalar of a large nearly spherical 2-surface embedded in $\Sigma $ \ was
proposed by Lau \cite{lau}. \ We shall consider the implications of using
Eq. (\ref{e7}) for Kerr black holes in spatially finite regions in the $%
l\rightarrow \infty $ limit. \ As in the asymptotically AdS case, although
there are terms of higher mass dimension that could be included, we shall
not consider these, since Eq. (\ref{e7}) is sufficient to remove divergences
at spatial infinity for this case. \ An algorithmic procedure, developed\
for asymptotically flat cases yields Eq. (\ref{e7}), although there are
topological restrictions on its applicability \cite{kls}.

A thorough discussion of the quasilocal formalism has been given elsewhere 
\cite{BY,BCM,ivan} and so we only briefly recapitulate it here. The action
can be written as a linear combination of \ a volume term Eq. (\ref{e3}), a
boundary term Eq. (\ref{e4}) and a counterterm\ which we shall take to be
Eq. (\ref{e7}) in either the small-$l$ or $l\rightarrow \infty $ limits:\ \
\ \ \ \ \ \ \ \ \ 
\begin{equation}
I=-\frac{1}{16\pi }\left\{ \int_{{\cal M}}d^{4}x\sqrt{-\gamma }{\cal R}%
+2\int_{\partial {\cal M}}d^{3}x\sqrt{-\gamma }\Theta -\int_{\partial {\cal M%
}}d^{3}x{\cal L}_{\mbox{ct}}\right\} .  \label{e8}
\end{equation}
\ Under the variation of the metric $\gamma ,$ one obtains: 
\begin{equation}
\delta I=[\mbox{terms that vanish when the equations of motion hold}]^{\mu
\nu }\delta g_{\mu \nu }+\int_{\partial {\cal M}}d^{3}x(P^{ij}+Q^{ij})]%
\delta \gamma _{ij}  \label{e9}
\end{equation}
where \ \ \ \ \ \ \ \ \ \ \ \ \ \ \ \ \ \ \ \ 
\begin{eqnarray}
P^{ij} &=&\frac{\sqrt{-\gamma }}{16\pi }(\Theta \gamma ^{ij}-\Theta ^{ij}),
\label{e10} \\
Q^{ij} &=&\frac{\sqrt{-\gamma }}{16\pi }\frac{\partial {\cal L}_{\mbox{ct}}}{%
\partial \gamma _{ij}},  \label{e11}
\end{eqnarray}
and Latin letters are used as indices for tensors on hypersurfaces.

Decomposing the metric $\gamma _{ij}$ on the timelike boundary $\partial
\Sigma _{\tau }\times \Upsilon ={\cal B}\times \Upsilon ={\cal T}$ which
connects the initial and final hypersurfaces 
\begin{equation}
\gamma _{ij}dx^{i}dx^{j}=-N^{2}dt^{2}+\sigma _{ab}\left( dx^{a}+V^{a}\right)
\left( dx^{b}+V^{b}\right)  \label{e12}
\end{equation}
yields after some manipulation 
\begin{equation}
\left. \delta I\right| _{{\cal T}}=\int_{{\cal T}}d^{3}x\sqrt{\sigma }\left(
-\varepsilon \delta N+j_{a}\delta V^{a}+\frac{1}{2}Ns^{ab}\delta \sigma
_{ab}\right) ,  \label{e13}
\end{equation}
where the coefficients of the varied fields are 
\begin{eqnarray}
\varepsilon &=&\frac{2}{N\sqrt{\sigma }}(P^{ij}+Q^{ij})u_{i}u_{j},
\label{e14} \\
j_{a} &=&-\frac{2}{N\sqrt{\sigma }}\sigma _{ai}(P^{ij}+Q^{ij})u_{j\mbox{ }},
\label{e15} \\
\ s^{ab} &=&\frac{2}{N\sqrt{\sigma }}\sigma _{\;i}^{a}\sigma
_{\;j}^{b}(P^{ij}+Q^{ij}),  \label{e16}
\end{eqnarray}
and we see that the counterterm variation $Q^{ij}$ supplants terms which
come from a reference action in the original formulation of the quasilocal
technique. \ The standard assumption \cite{BY} for the reference action is
that it is a linear functional of the lapse $N$ and shift $V^{a}$ so that
its contributions to $\varepsilon $ and $j_{a}$\ depend only on the induced
two-metric $\sigma _{ab}$, whose determinant we denote by $\sigma $. This
assumption does not hold for the counterterm prescription (\ref{e7}). \ \ 

Since $-\sqrt{\sigma }\varepsilon $ is the time rate of change of the
action, $\varepsilon $ is identified with the energy density on the surface $%
{\cal B}$, and the total quasilocal energy for the system is therefore 
\begin{equation}
E=\int_{{\cal B}}d^{2}x\sqrt{\sigma }\varepsilon ,  \label{e17}
\end{equation}
and this quantity can be meaningfully associated with the thermodynamic
energy of the system \cite{jolthesis}.Using similar reasoning, the
quantities $j_{a}$ and $s^{ab}$ are respectively referred to as the momentum
surface density and the spatial stress.

When there is a Killing vector field ${\cal \xi }$\ on the boundary ${\cal T}
$ , an associated conserved charge is defined by 
\begin{equation}
{\cal Q}\left( {\cal \xi }\right) =\int_{{\cal B}}d^{2}x\sqrt{\sigma }\left(
\varepsilon u^{i}+j^{i}\right) {\cal \xi }_{i}  \label{e18}
\end{equation}
provided there is no matter stress energy in the neighborhood of ${\cal T}$
\ (this assumption can be dropped, allowing one to compute the time rate of
change of ${\cal Q}$, if desired \cite{jolivan}). \ When this holds, the
value of ${\cal Q}$ is independent of the particular hypersurface ${\cal B}$
, a property not shared by the energy $E$. \ For boundaries with timelike ($%
\xi =\partial /\partial t$) and rotational ($\varsigma =\partial /\partial
\phi $) Killing vector fields \ (which encompass all the metrics we consider
in this paper) we obtain 
\begin{eqnarray}
M &=&\int_{{\cal B}}d^{2}x\sqrt{\sigma }\left( \varepsilon
u^{i}+j^{i}\right) \xi _{i}  \label{e19} \\
J &=&\int_{{\cal B}}d^{2}x\sqrt{\sigma }j^{i}\varsigma _{i}  \label{e20}
\end{eqnarray}
provided the surface ${\cal B}$ contains the orbits of $\varsigma $. These
quantities are respectively the conserved mass and angular momentum of the
system enclosed by the boundary. Note that they will both be dependent upon
the location of the boundary ${\cal B}$ in the spacetime, although each is
independent of the particular choice of foliation ${\cal B}$ within the
surface ${\cal T}$. \ 

In the context of the AdS/CFT correspondence the limit in which the boundary 
${\cal B}$ becomes infinite is taken, and the counterterm prescription \cite
{hensken,hyun,balakraus,kls} ensures that the conserved charges (\ref{e18})
are finite. No embedding of the surface ${\cal T}$ \ into a reference
spacetime is required. This is of particular advantage for the class of Kerr
spacetimes, in which it is not possible to embed an arbitrary
two-dimensional boundary surface into a flat (or constant-curvature)
spacetime \cite{Martinez}. The counterterm (\ref{e7}) includes the minimal
number of boundary curvature invariants for which this holds in both the
large-$l$ and ${\cal O}\left( l\right) $ limits. \ We shall consider the
extension of Eq. (\ref{e7}) away from the ${\cal B}\rightarrow \infty $
limit in what follows.

The class of metrics we shall henceforth consider are the Kerr-AdS family of
solutions, whose general form is 
\begin{eqnarray}
ds^{2} &=&-\frac{\Delta _{r}}{\rho ^{2}}(dt-\frac{a}{\Xi }\sin ^{2}\theta
d\phi )^{2}+\frac{\rho ^{2}}{\Delta _{r}}dr^{2}+\frac{\rho ^{2}}{\Delta
_{\theta }}d\theta ^{2}  \label{kadsmet} \\
&&+\frac{\Delta _{\theta }\sin ^{2}\theta }{\rho ^{2}}[adt-\frac{%
(r^{2}+a^{2})}{\Xi }d\phi ]^{2}  \nonumber
\end{eqnarray}
in $(3+1)$ dimensions, where 
\begin{eqnarray}
\Delta _{r} &=&(r^{2}+a^{2})(1+r^{2}/l^{2})-2mr;  \nonumber \\
\Delta _{\theta } &=&1-a^{2}\cos ^{2}\theta /l^{2};  \label{kadsnotn} \\
\Xi &=&1-a^{2}/l^{2};  \nonumber \\
\rho ^{2} &=&r^{2}+a^{2}\cos ^{2}\theta ,  \nonumber
\end{eqnarray}
and where $\Lambda =-3/l^{2}$. \ The metric has two horizons located at $%
r_{\pm }$ ,\ provided the parameter $m$ is sufficiently large relative to
the other parameters, specifically 
\begin{equation}
m\geq \frac{l}{3\sqrt{6}}\left( \sqrt{1+14\frac{a^{2}}{l^{2}}+\frac{a^{4}}{%
l^{4}}}+2\left( 1+\frac{a^{2}}{l^{2}}\right) \right) \sqrt{\sqrt{1+14\frac{%
a^{2}}{l^{2}}+\frac{a^{4}}{l^{4}}}-\left( 1+\frac{a^{2}}{l^{2}}\right) }%
\equiv m_{\mbox{crit}}  \label{kadscrit}
\end{equation}
In the limit $l\rightarrow \infty $, $r_{\pm }=m\pm \sqrt{m^{2}-a^{2}}$, and 
$m_{\mbox{crit}}=a$.

For $m=0,$ the metric (\ref{kadsmet}) is that of pure AdS spacetime (or flat
spacetime if $l\rightarrow \infty $), and for $a=0,$ the metric is that of
Schwarzchild-AdS spacetime which has zero angular momentum. \ Hence we
expect the parameters $a$ and $m$ to be associated with the mass and angular
momentum of the spacetime, respectively. This can be confirmed using
conformal completion methods \cite{ashdas}, for which we find that the total
mass $M_{t}$ and angular momentum $J_{t}$ are given by \cite{dasmann}

\begin{equation}
M_{t}=\frac{m}{\Xi }\mbox{ \ \ \ \ \ \ \ \ },\quad J_{t}=\frac{am}{\Xi
^{2}},  \label{kadsmj}
\end{equation}
a result corroborated by counterterm methods \cite{nutkerr}.

We turn now to consider the extension of (\ref{e7}) away from the ${\cal B}%
\rightarrow \infty $ limit.

\section{Kerr Metric}

\bigskip

Here we take $l\rightarrow \infty $, and so the action can be written as 
\begin{equation}
I=-\frac{1}{16\pi }\left\{ \int_{{\cal M}}d^{4}x\sqrt{-\gamma }{\cal R}%
+2\int_{\partial {\cal M}}d^{3}x\sqrt{-\gamma }\Theta -2\sqrt{2}%
\int_{\partial {\cal M}}d^{3}x\sqrt{-\gamma }\sqrt{{\cal R}(\gamma )}\right\}
\label{km1}
\end{equation}
\ We write, for convenience, $Q^{ij}=Q_{2}^{ij}+Q_{3}^{ij}$ where 
\begin{eqnarray}
Q_{2}^{ij} &=&\frac{\sqrt{-\gamma }}{16\pi }\sqrt{\frac{2}{{\cal R}}}(R^{ij}-%
{\cal R}\gamma ^{ij})  \label{km4} \\
Q_{3}^{ij} &=&\frac{\sqrt{-\gamma }}{16\pi }\frac{1}{\sqrt{2}}\left( \nabla
_{a}(\nabla ^{a}\,{\cal R}^{-1/2})\gamma ^{ij}-\frac{1}{2}\nabla
^{(i}(\nabla ^{j)}\,{\cal R}^{-1/2}\,)\right)  \label{km5}
\end{eqnarray}
and the energy of the system can be written as : 
\begin{equation}
E=E_{1}+E_{2}+E_{3};\ \ \ \ \ \ \ \ \ \ \ \ \ \ \ \ \ \ E_{i}=\int_{{\cal B}%
}d^{2}x\sqrt{\sigma }\varepsilon _{i},  \label{km6}
\end{equation}
where ${\cal B}$ is a 2-dimensional surface defined by setting the radial
coordinate to a constant value \ $r=R$, and $\varepsilon _{1},$ $\varepsilon
_{2},$ $\varepsilon _{3}$ are given by \ 
\begin{eqnarray}
\varepsilon _{1} &=&\frac{2}{N\sqrt{\sigma }}P^{ij}u_{i}u_{j}  \label{km7} \\
&=&\frac{1}{4}R\frac{\sqrt{1+A^{2}-2M}[(M-1)A^{2}z^{2}-2-(1+M)A^{2}]}{%
\{(1+A^{2}z^{2})[(1+A^{2}-2M)A^{2}z^{2}+1+A^{2}+2MA^{2}]\}^{1/2}}  \nonumber
\\
\varepsilon _{2} &=&\frac{2}{N\sqrt{\sigma }}Q_{2}^{ij}u_{i}u_{j\mbox{ }}
\label{km8} \\
&=&\frac{1}{2}R\{(A^{4}-2M)A^{4}z^{4}+(2A^{2}-7MA^{2}+2-4MA^{4}-7M)A^{2}z^{2}
\nonumber \\
&&+1+A^{2}+3MA^{2}+A^{4}M\}  \nonumber \\
&&\times
\{(1-2MA^{2}z^{2}+A^{2}z^{2})[(1+A^{2}-2M)A^{2}z^{2}+1+2MA^{2}+A^{2}](1+A^{2}z^{2})^{3}\}^{-1/2}
\nonumber \\
\ \varepsilon _{3} &=&\frac{2}{N\sqrt{\sigma }}Q_{3}^{ij}u_{i}u_{j}
\label{km9} \\
&=&\frac{1}{2}R\{(1-2M^{2})(1+A^{2}-2M)A^{8}z^{10}  \nonumber \\
&&+(1-2M)(12M^{2}-4A^{2}M^{2}-24M-6MA^{2}+9+9A^{2})A^{6}z^{8}  \nonumber \\
&&+\left[ (2M-1)^{3}A^{4}-\left( 36M^{3}-42M^{2}-7M+14\right)
A^{2}+(2M-1)\left( 2M^{2}+15M-15\right) \right] A^{4}z^{6}  \nonumber \\
&&+\left[ -3(2M-1)(M-1)A^{4}-\left( 8M^{3}+36M^{2}-25M-8\right)
A^{2}-8M^{2}-12M+11\right] A^{2}z^{4}  \nonumber \\
&&+(3-8A^{4}M^{3}-3A^{4}+3M+15A^{2}M+4A^{4}M^{2}+6A^{2}M^{2})z^{2}  \nonumber
\\
&&-1-M-A^{2}-2A^{2}M^{2}-3A^{2}M\}  \nonumber \\
&&\times
\{(1-2MA^{2}z^{2}+A^{2}z^{2})^{5}[(1+A^{2}-2M)A^{2}z^{2}+1+2MA^{2}+A^{2}](1+A^{2}z^{2})^{3}\}^{-1/2}
\nonumber
\end{eqnarray}
where $z=\cos \theta $ and $A=a/R$ and $M=m/R$.

For arbitrary values of $A\ $and $M$, the term $E_{1}$ can be integrated
analytically in terms of Elliptic functions. Although the terms $E_{2%
\mbox{
\ \ }}$and $\ E_{3}$ cannot be analytically integrated, a term-by-term
expansion of $E_{3}$ in powers of $A$ indicates that it is zero, a fact that
can also be verified \ numerically.

\bigskip

For small angular momentum $A$, $E$ can be integrated easily: 
\begin{equation}
E=R\{1-\sqrt{1+A^{2}-2M}+\frac{A^{2}}{6}[2(1+M)+(1+2M)\sqrt{1+A^{2}-2M}%
]+O(A^{4})]\}.  \label{km10}
\end{equation}
In the asymptotic limit $R\rightarrow \infty ,$the energy approaches the
Arnowitt-Deser-Misner (ADM) energy $M$, \ \ and for the case of $A=0,$ one
gets the thermodynamic energy of a Schwarzschild black hole, which is
equivalent to its mass. \ The expression (\ref{km10}) was obtained by
Martinez \cite{Martinez} using the quasilocal method with\ a flat reference
spacetime in which the constant-$R$ boundary was embedded in the small-$A$
limit. However even in this limit the embedding was technically quite
complicated. We find it interesting that the counterterm prescription (\ref
{km1}), extended to finite values of fixed $R,$ gives exactly the same
answer to this order in $A$.

\bigskip

We next consider the total energy $E$ in (\ref{km6}) for $R=r_{+}$ where $%
r_{+}$ is the radius of \ the outer horizon. In this case we can exactly
compute $E$ in terms of elliptic functions: 
\begin{equation}
E(r_{+})=-\frac{1+A_{+}^{2}}{2}r_{+}[(4+\frac{3}{A_{+}^{2}})F(A_{+}^{2},%
\frac{i}{A_{+}})-5E(A_{+}^{2},\frac{i}{A_{+}})-5\sqrt{1-A_{+}^{2}}]
\label{kmexact}
\end{equation}
where $i^{2}=-1$, $A_{+}=a/r_{+}$, and 
\[
E\left( z,k\right) =\int_{0}^{z}dt\frac{\sqrt{1-k^{2}t^{2}}}{\sqrt{1-t^{2}}}%
\qquad F\left( z,k\right) =\int_{0}^{z}dt\frac{1}{\sqrt{1-t^{2}}\sqrt{%
1-k^{2}t^{2}}} 
\]
are the definitions of the elliptic functions $E(z,k)$ and $F(z,k)$ .

\bigskip

\begin{figure}[tbp]
\begin{center}
\epsfig{file=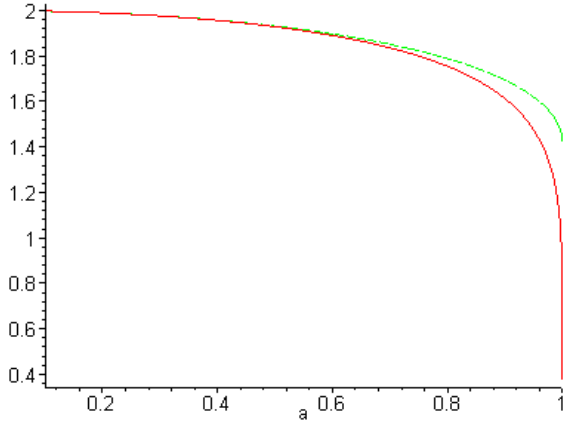,width=0.8\linewidth}
\end{center}
\caption{$E\left( r_{+}\right) /m$ (solid) , and $2M_{i}/m$ (dashed) \
versus $a$ plotted in units of $m$.}
\label{Figure1}
\end{figure}

\begin{figure}[tbp]
\begin{center}
\epsfig{file=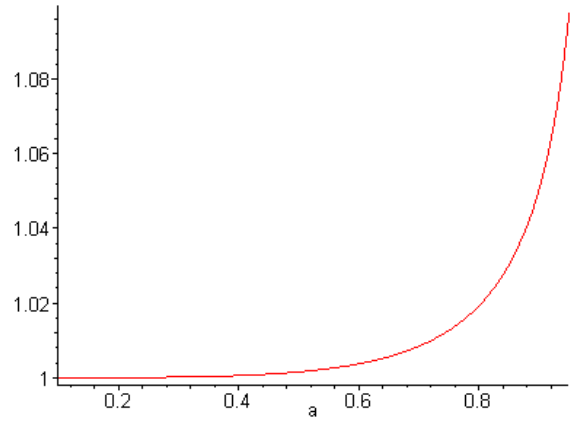,width=0.8\linewidth}
\end{center}
\caption{$M_{i}/E(r_{+})$ versus $a$ plotted in units of $m$.}
\label{Figure2}
\end{figure}

In Fig.. \ref{Figure1}\ we plot the $A_{+}$-dependence of the energy at the
horizon. In the slow-rotation approximation it has been shown that 
\begin{equation}
E\left( r_{+}\right) =2M_{i}\left[ 1+{\cal O}\left( \frac{a^{4}}{M_{i}^{4}}%
\right) \right] ,  \label{km10a}
\end{equation}
where 
\begin{equation}
M_{i}=\frac{1}{2}\sqrt{r_{+}^{2}+a^{2}}  \label{km10b}
\end{equation}
is the irreducible mass of the Kerr black hole \cite{Christodolou}, the
maximal amount of energy which can be extracted from a black hole via the
Penrose process. It was conjectured in Ref. \cite{Martinez} that the
relation $E\left( r_{+}\right) =2M_{i}$ is an exact relation, valid for any
value of $A$. We see using the counterterm prescription that this conjecture
is not correct and that (\ref{km10a}) is only an approximation which breaks
down for $\frac{a}{R}>0.5$, as shown in Figs. \ref{Figure1} and \ref{Figure2}%
.

\bigskip 

\bigskip

\bigskip 

We next consider evaluation of the energy for fixed arbitrary values of $R$.
The calculation of $E_{2\mbox{ }}$ must be carried out numerically, and in a
manner which yields a well defined integral. This means that the parameters
of the integral must be such that the location of the quasilocal surface
must be outside of the event horizon and the induced Ricci scalar ${\cal R}%
(\gamma )$ must be positive. These criteria are satisfied provided that $A$
and $M$ are such that : 
\begin{eqnarray}
&&\mbox{a) For }M\leq 0.5\mbox{ then}\ 0\leq A\leq M,\mbox{\ }  \nonumber \\
&&\mbox{b) For }0.5\leq M\leq 1\mbox{ then }\sqrt{2M-1}\leq A\leq M.
\label{kerrconds} \\
&&\mbox{c) }A\leq 1,\ M\leq 1  \nonumber
\end{eqnarray}

We plot the dependence of $E(M,A)$ in Figs. \ref{Figure3} and \ref{Figure 4}
below, consistent with these constraints. As one can see from Fig. \ref
{Figure3}, for small values of \ $M$ ($M<0.4$) the energy is a weakly
varying function of $\ A$ but as $M$ increases the $A$-dependence of energy
becomes more relevant. For example the average value of $\left| \partial
E/\partial A\right| $ is approximately $10^{-2}$ and $10$ for $M=0.1$ and $%
M=0.8$ respectively. Of course these values of $M=m/R$ depend on the value
of the radius $R$ which is chosen for the quasilocal surface.

\bigskip

\begin{figure}[tbp]
\begin{center}
\epsfig{file=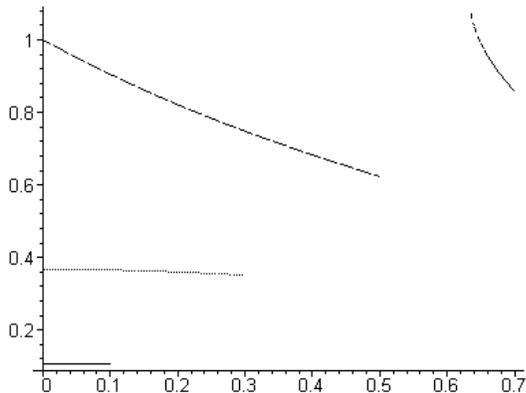,width=0.8\linewidth}
\end{center}
\caption{$E/R$ versus $A$ for $M=0.1$ (solid), $0.3$ (dotted), $0.5$
(dashed), $0.7$ (dot-dashed).}
\label{Figure3}
\end{figure}

\begin{figure}[tbp]
\begin{center}
\epsfig{file=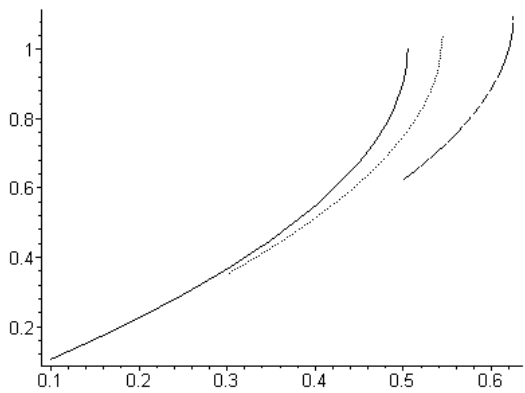,width=0.8\linewidth}
\end{center}
\caption{$E/R$ versus $M$ for $A=0.1$ (solid), $0.3$ (dotted), $0.5$
(dashed).}
\label{Figure 4}
\end{figure}

\bigskip

Now we consider the quasilocal conserved charges of the Kerr metric. The
Kerr metric has two Killing vectors and therefore there exist two conserved
quantities. The first one, which is due to the timelike Killing vector $\xi
^{a}=(\partial /\partial t)^{a}$, is associated with the conserved mass, $%
{\cal M},$ of the system expressed as: 
\begin{equation}
{\cal M}=-\int_{B}d^{2}x\sqrt{\sigma }\frac{2}{N}%
(P^{ij}+Q_{2}^{ij}+Q_{3}^{ij})u_{i}\xi _{j},  \label{km12}
\end{equation}
where $N$ is the lapse function. For the case of \ $R=r_{+}$, we can
explicitly compute the conserved mass: 
\begin{equation}
{\cal M}\left( r_{+}\right) = \frac{1}{2}r_{+}A^{2}=\frac{a^{2}} {2r_{+}}=m-%
\frac{1}{2}r_{+}=\frac{1}{2}r_{-}  \label{km13}
\end{equation}
which should be compared with the expression for the quasilocal energy. In
the limit of slow rotation it vanishes, a strikingly different feature from
that of the energy as given in eq. (\ref{km10a}) , which in this limit
equals $2m$ . As $R\rightarrow \infty $ we find that $M\rightarrow m$ as
expected.

\begin{figure}[tbp]
\begin{center}
\epsfig{file=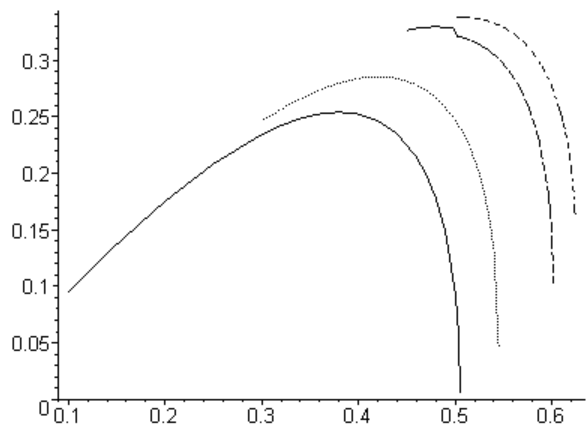,width=0.8\linewidth}
\end{center}
\caption{${\cal M}/R$ versus $M$ for $A=.1$ (solid), $.3$ (dotted), $0.45$
(dashed), $0.5$\ (dot-dashed).}
\label{Figure 5}
\end{figure}

\begin{figure}[tbp]
\begin{center}
\epsfig{file=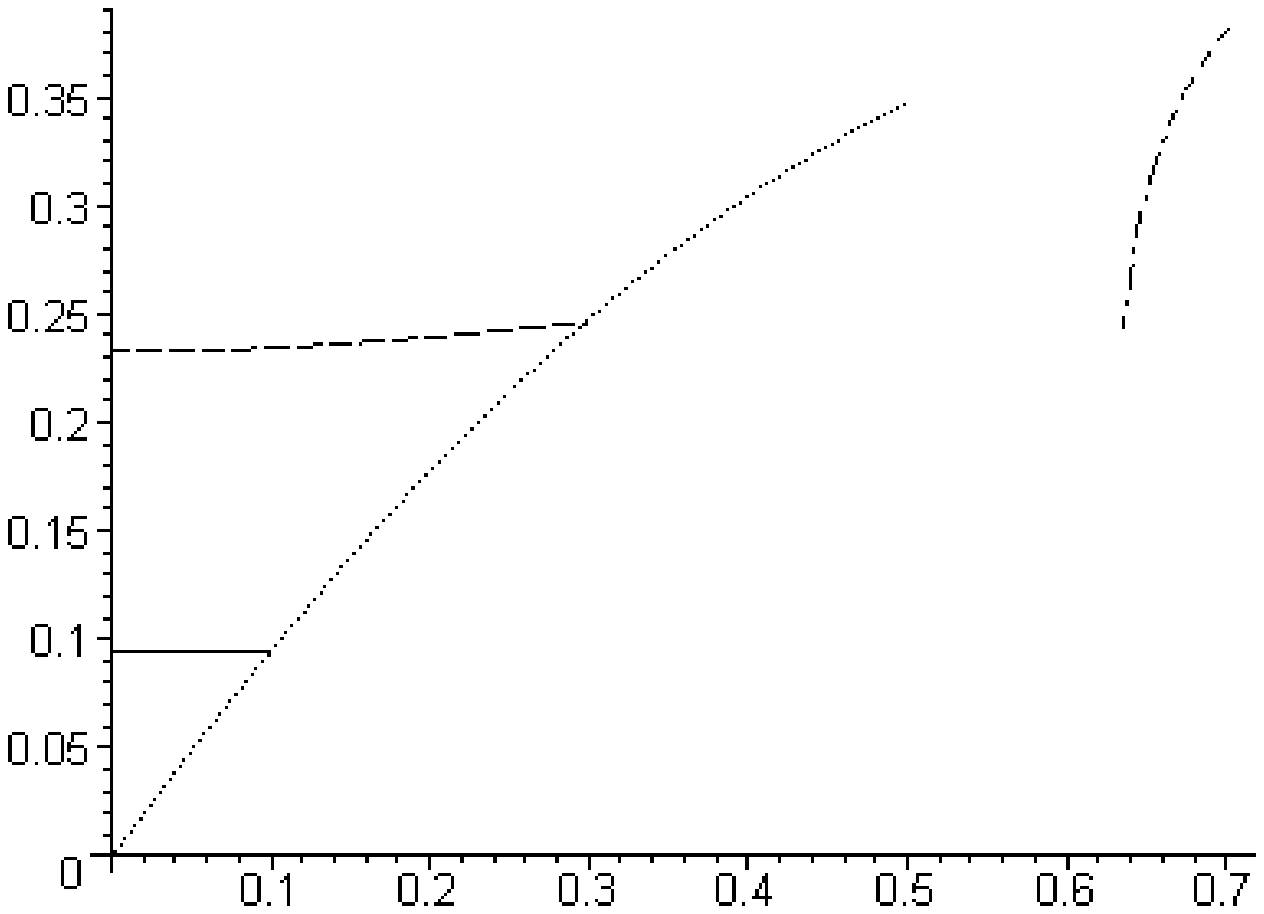,width=0.8\linewidth}
\end{center}
\caption{$\frac{{\cal M}}{R}$ versus $A$ for $M=.1$ (solid), $0.3$ (dotted), 
$0.5$ (dashed), $0.7$\ (dot-dashed).}
\label{Figure6}
\end{figure}

It is clear from Figs. \ref{Figure3} and \ref{Figure 4} that the energy for
a fixed value of $A$ is an increasing function of $M.$ This is not the case
for the conserved mass ${\cal M}(M,A)$ in (\ref{km12}), which for fixed $%
A<0.5$ has a maximum value for $M\geq \frac{1}{3}$, with the bound being
saturated for the Schwarzchild solution when $A=0$. We plot in Figs. \ref
{Figure 5} and \ref{Figure6} the behavior of the conserved charge ${\cal M}%
\left( M,A\right) $ for fixed $A$ and $M$ respectively. We find that the
value of ${\cal M}$ due to the $Q_{3}^{ij}$ third term in eq. (\ref{km12})
is very small compared to that of the first two terms.

\bigskip

\bigskip

The second non-trivial conserved charge is the $\varphi $ component of the
angular momentum, $J_{\varphi },$ which can be expressed as : 
\begin{equation}
J_{\varphi }=-\int d^{2}x\sqrt{\sigma }\frac{2}{N}%
(P^{ij}+Q_{2}^{ij}+Q_{3}^{ij})u_{i}\sigma _{ja}\zeta ^{a},  \label{km14}
\end{equation}
where \ $\zeta ^{a}$ is Killing vector $(\frac{\partial }{\partial \varphi }%
).$ Remarkably we find that the counter terms $Q_{2}^{ij}+Q_{3}^{ij}$ ,
while non-zero, give a vanishing contribution upon integration in (\ref{km14}%
). Hence the total angular momentum is due entirely to the first term, which
is:

\begin{equation}
J_{\varphi }=r^{2}MA=ma  \label{km15}
\end{equation}
valid for any quasilocal surface of fixed $r=R$.

We turn next to thermodynamic considerations of the Kerr black hole. The
interior of the quasilocal surface can be regarded as a thermodynamic system
which is co-rotating with an exterior heat bath whose angular velocity is $%
\Omega _{H}$ \cite{BMY}, where $\Omega _{H}=\frac{a}{r_{+}^{2}+a^{2}}$ is
the angular velocity of the event horizon. \ Zero angular momentum observors
(ZAMO's) at a given position outside of the black hole measure a proper
inverse temperture $\beta _{\mbox{ZAMO}}=N\beta $ and a proper angular
velocity $\Omega _{\mbox{ZAMO}}\left( r,\theta \right) =N^{-1}\left( \Omega
_{H}-\frac{d\phi }{dt}\right) $. \ However quasilocal thermodynamic
quantities are generally given by surface integrals over the boundary data
rather than by products of constant surface data. In particular, it is not
possible to choose the quasilocal surface to simultaneously be both
isothermal and a surface of constant $\Omega _{\mbox{ZAMO}}$ and neither is
true for the quasilocal surface we have chosen (one at fixed $r=R$). \
Furthermore it is by now well recognized that the entropy of a black hole
depends only on the geometry of its horizon in the classical approximation,
and is conversely independent of the asymptotic behavior of the
gravitational field or of the presence of external matter fields. \ For
stationary black holes, the entropy is one-quarter of the horizon area,
yielding\ $S=\pi \left( r_{+}^{2}+a^{2}\right) =2\pi R^{2}M(M+\sqrt{%
M^{2}-A^{2}})$ in this case. We therefore shall regard the energy $E$ in Eq.
(\ref{km6}) as the thermodynamic internal energy for the Kerr black hole
within the boundary $r=R$, view it as a function of \ $R$ , the entropy $S$,
and the angular momentum $J$, and define the temperature and the angular
velocity from their ensemble derivatives, integrated over the quasilocal
boundary data.

For the temperature we obtain

\begin{equation}
T=\left( \frac{\partial E}{\partial S}\right) _{R,J}=\frac{1}{4\pi R}\frac{%
\sqrt{M^{2}-A^{2}}}{M(M+\sqrt{M^{2}-A^{2}})}\left( \frac{\partial E}{%
\partial M}\right) _{R,J}=\frac{\kappa _{+}}{2\pi R}\left( \frac{\partial E}{%
\partial M}\right) _{R,J}  \label{km11}
\end{equation}
where $\kappa _{+}$\ is the surface gravity at the outer horizon. The
derivative $\left( \frac{\partial E}{\partial M}\right) _{R,J}$ is obtained
by taking the derivatives of the terms in Eqs. (\ref{km7} --\ref{km9} )
above and then performing the integration with respect to $z=\cos \theta $.
It is straightforward to show that ($\partial E/\partial M)_{R,J}\rightarrow
1$ as $R\rightarrow \infty $, yielding the usual relationship between
temperature and surface gravity. However unlike the non-rotating case, it
does not give the Tolman redshift factor because a surface of fixed $r=R$ is
not a surface of constant redshift (it is not an isotherm). \ 

\begin{figure}[tbp]
\begin{center}
\epsfig{file=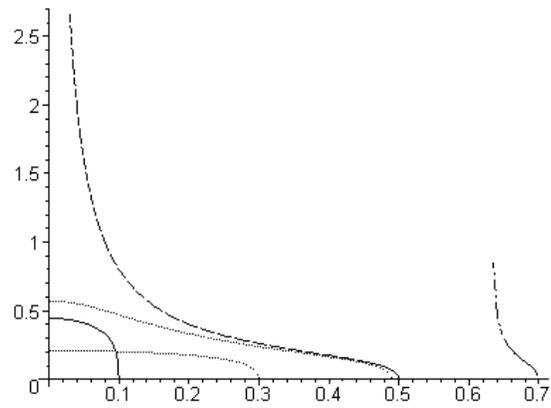,width=0.8\linewidth}
\end{center}
\caption{$RT$ versus $A$ for $M=0.1$ (solid), $0.3$ (dotted), $0.49$ (light
dotted) $0.5$ (dashed), $.7$ (dot-dashed).}
\label{Figure7}
\end{figure}

\begin{figure}[tbp]
\begin{center}
\epsfig{file=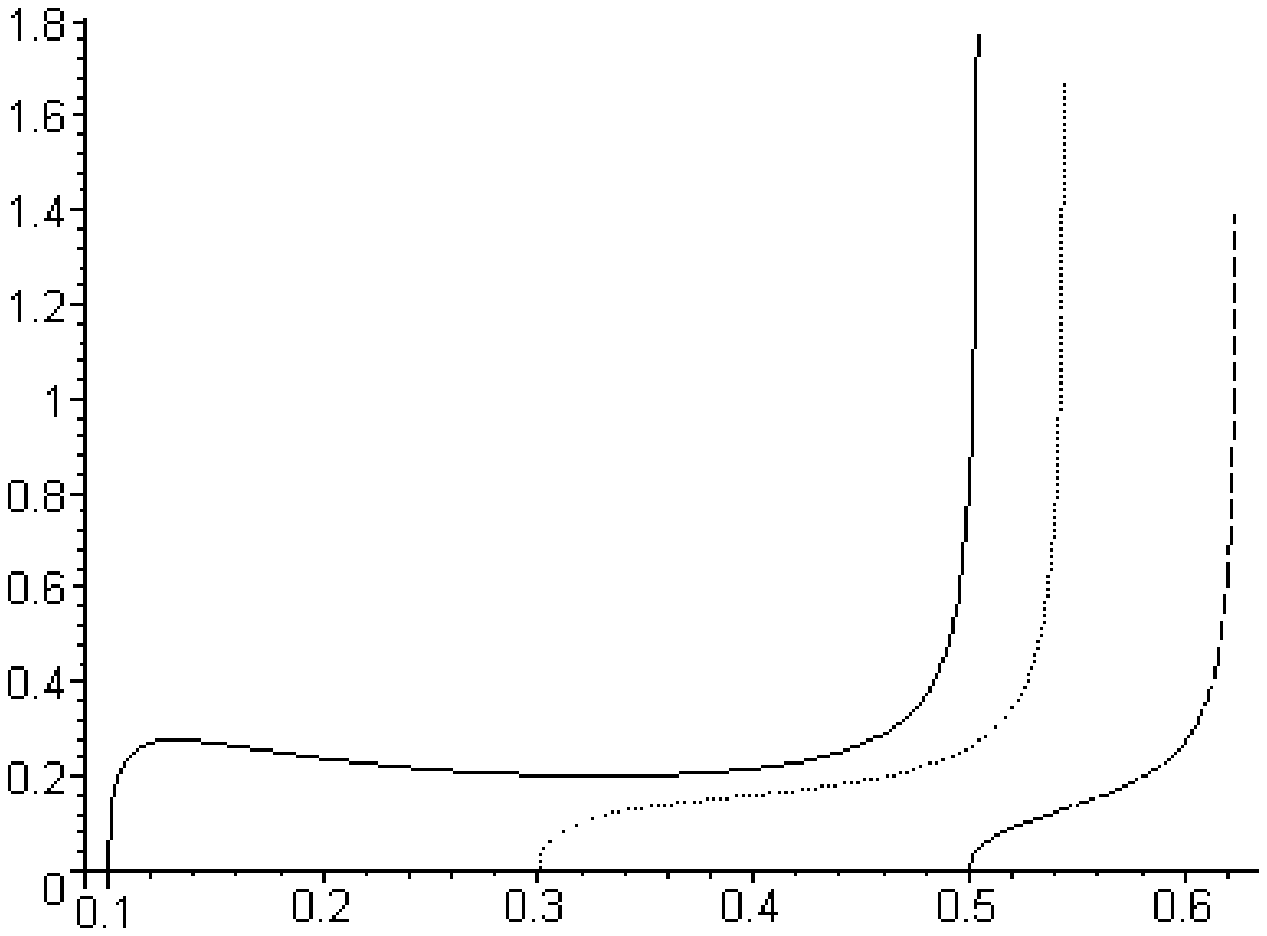,width=0.8\linewidth}
\end{center}
\caption{$RT$ versus $M$ for $A=0.1$ (solid), $0.3$ (dotted), $0.5$\
(dashed).}
\label{Figure8}
\end{figure}

\begin{figure}[tbp]
\begin{center}
\epsfig{file=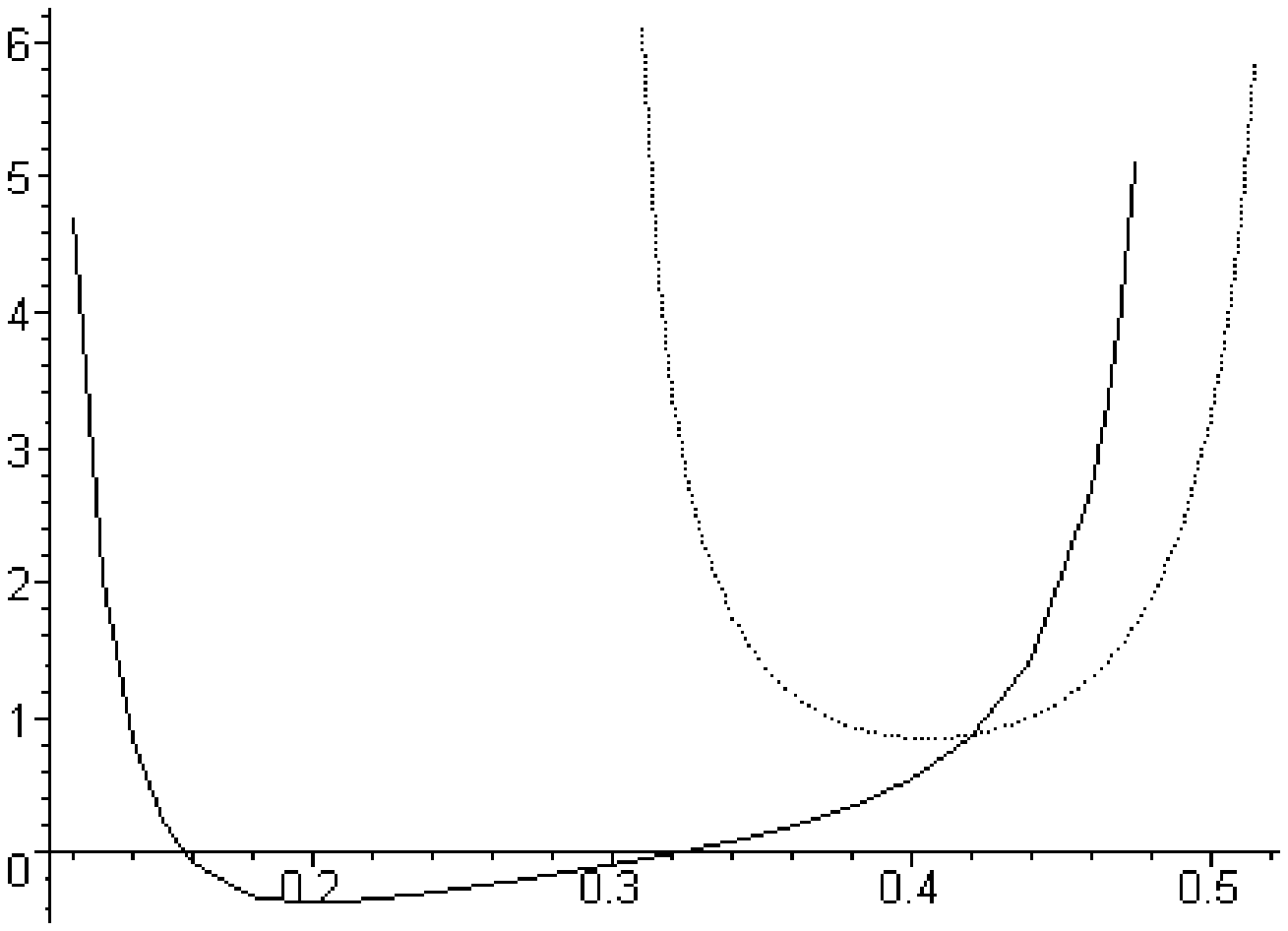,width=0.8\linewidth}
\end{center}
\caption{$R(\partial T/\partial M)$ versus $M$ for $A=0.1$ (solid), $0.3$
(dotted).}
\label{Figure9}
\end{figure}

Now we investigate the $A$ and $M$ dependence of the temperature of a Kerr
black hole. The $A$ dependence of the temperature is shown in Fig. \ref
{Figure7}. For $M<0.4$ the temperature does not depend on $A$ very much
(except for $(M-A)/M\ll 1,$where $T\longrightarrow 0$ as $A\rightarrow M$),
but as $M$ increases ( $M\geq 0.5$) the $A$-dependence of $T$ becomes more
relevant. For the case of $M>0.5$ as $r\rightarrow r_{+}$ then $T$ goes to
infinity and it goes to zero as $A$ goes to $M$. The $M$-dependence of $T$\
is shown in fig. \ref{Figure8}. The $A$-dependence of $(\frac{\partial T}{%
\partial M})_{J,R}$ can be seen from fig. \ref{Figure9}.

\bigskip

\bigskip

The temperature for a fixed value of $J$ does not increase monotically. For
example from fig. \ref{Figure9} for $A=0.1,$ it can be seen that $(\partial
T/\partial M)_{J,R}$ is zero at $M\simeq 0.16$, and $M\simeq 0.32.$ These
two points correspond to the maximum and minimum of $T(M,J,R)$ respectively.
\ $T(M,J,R)$ for the Schwarzschild case with $A=0,$ has its minimum at $%
M=1/3.$ For small values of $A$, the situation is similar; at a fixed value
of the temperature $T$ below $T_{\mbox{min}}$ there will be only one black
hole solution of small $M$, in equilibrium with thermal radiation inside the
box of characteristic radius $r=R$, which will not be far from the extremal
solution. \ At $T$ increases above $T_{\mbox{min}}$ two more possible
solutions appear, one of slightly larger $M$ and one of much larger $M$. \
As $T$ increases above $T_{\mbox{max}}$ the two smaller $M$ coalesce,
leaving only the largest $M$ solution. As $A$ increases the region for the
three distinct black hole solutions becomes vanishingly small, and for
sufficiently large $A,$ there will be only one black hole solution of large $%
M$.

\bigskip

The heat capacity at constant surface area, $4\pi R,$ is defined by: 
\begin{equation}
C_{R}=\left( \frac{\partial E}{\partial T}\right) _{R,J},  \label{km16}
\end{equation}
where the energy is expressed as a function of $T$ and $R.$ Expressing the
energy and temperature as functions of $M,J$ and $R,$ then the heat capacity
can be written as

\begin{equation}
C_{R}=\left( \frac{\partial E}{\partial M}\right) _{R,J}\left( \frac{%
\partial T}{\partial M}\right) _{R,J}^{-1}.  \label{km17}
\end{equation}

\begin{figure}[tbp]
\begin{center}
\epsfig{file=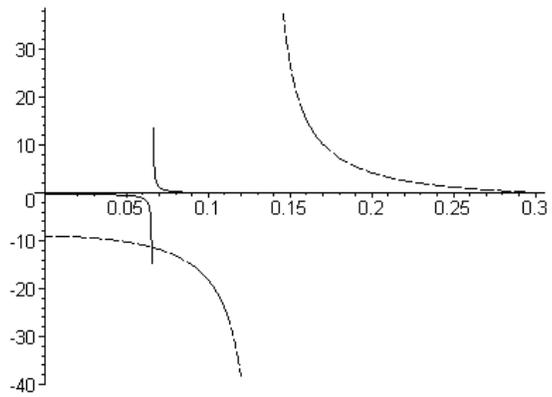,width=0.8\linewidth}
\end{center}
\caption{$C_{R}/R^{2}$ versus $A$ for $M=0.1$ (solid), $0.3$ (dashed).}
\label{Fig13}
\end{figure}

For small values of $M$ ($M<1/3$), the heat capacity is not positive for all
the allowed values of $A.$ For example as one can see from fig. \ref{Fig13}
for $M=0.1,$ the heat capacity is positive only for $A>0.055$. Indeed, for
each value of $M$ $<1/3$ there exists an $A$ for which the heat capacity
becomes infinite . For $M\approx 1/3$ the heat capacity is negative for a
Schwarzschild black hole. However for a Kerr black hole we see from fig. \ref
{Figure 11} there are some values of $A$ for which the heat capacity is
positive. There is a phase of stable small black holes and a phase of stable
large black holes, separated by a regime of intermediate unstable black
holes, as illustrated in fig. \ref{Figure9}.

\begin{figure}[tbp]
\begin{center}
\epsfig{file=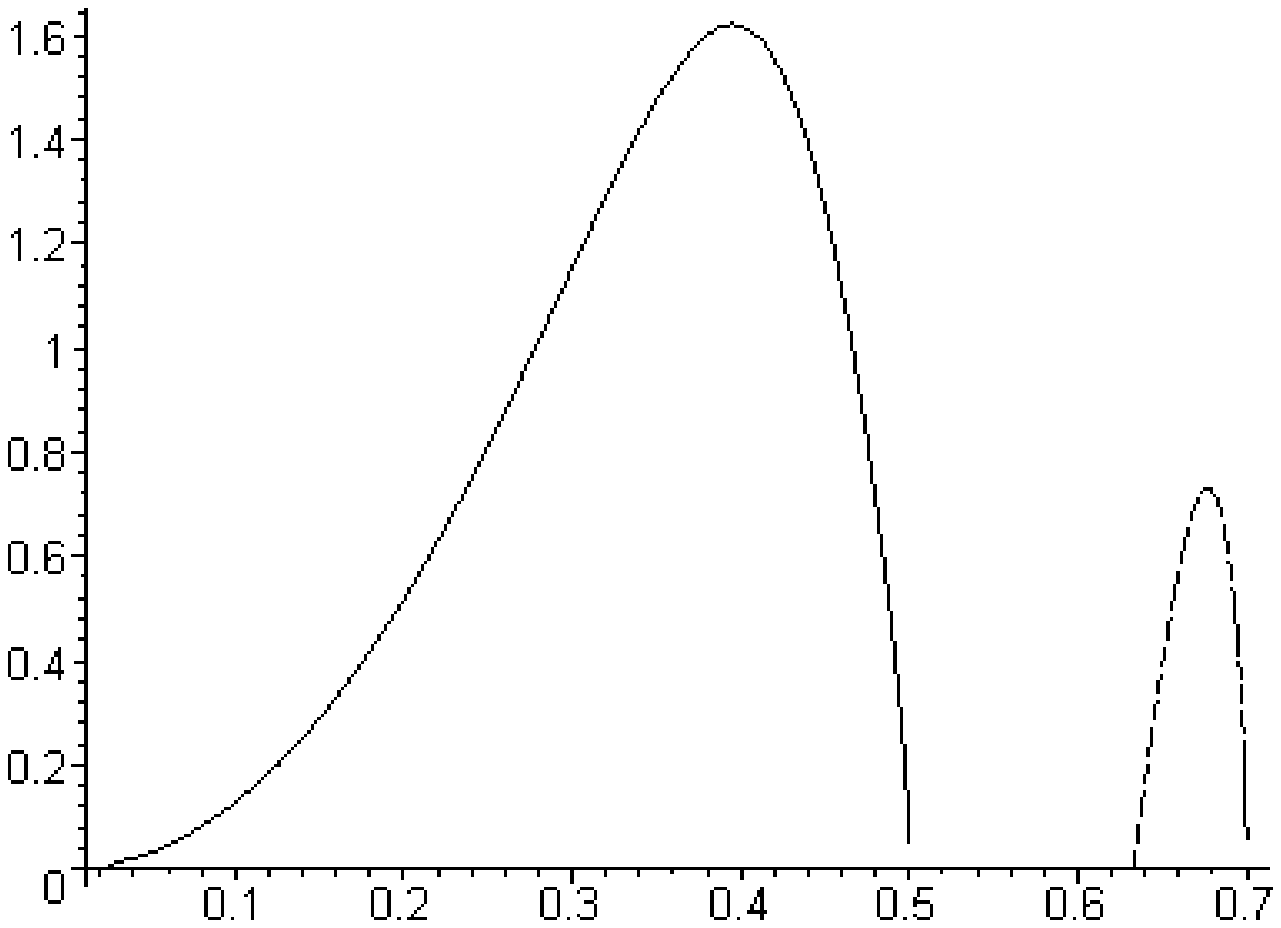,width=0.8\linewidth}
\end{center}
\caption{$C_{R}/R^{2}$ versus $A$ for $M=0.5$ (solid), $0.7$ (dashed).}
\label{Figure11}
\end{figure}

This fact was noted long ago by Davies, who gave the relation 
\begin{equation}
a^{4}+6a^{2}m^{2}-3m^{4}=0
\end{equation}
for the values of $a/m$ at which the heat capacity becomes infinite \cite
{Davies} in the limit $R\rightarrow \infty $. Here since we use the
quasilocal energy at $r=R$ instead of the ADM mass parameters the
relationship is considerably more complicated, and can only be obtained
numerically. However the qualitative features remain the same. The $A$%
-dependence of the heat capacity for larger values of $M$ can be seen from
Figs. \ref{Fig13} and \ref{Figure11}. For sufficiently large $M$, the heat
capacity is positive for all allowed values of $A$, and there is a single
phase consisting of a large black hole.

\bigskip

\bigskip

\bigskip

\begin{figure}[tbp]
\begin{center}
\epsfig{file=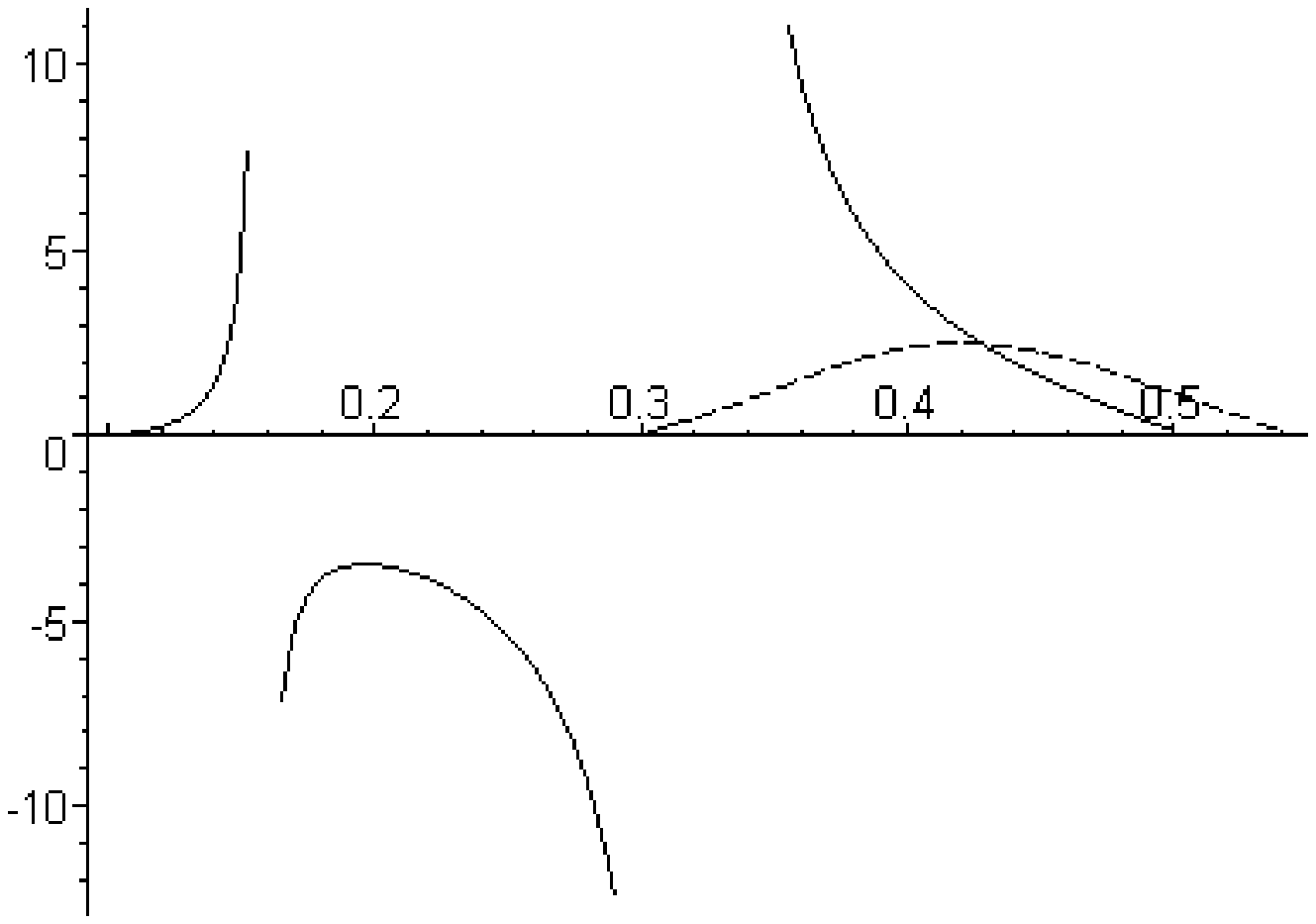,width=0.8\linewidth}
\end{center}
\caption{$C_{R}/R^{2}$ versus $M$ for $A=0.1$ (solid), $0.3$ (dashed).}
\label{Figure12}
\end{figure}

\ 

\begin{figure}[tbp]
\begin{center}
\epsfig{file=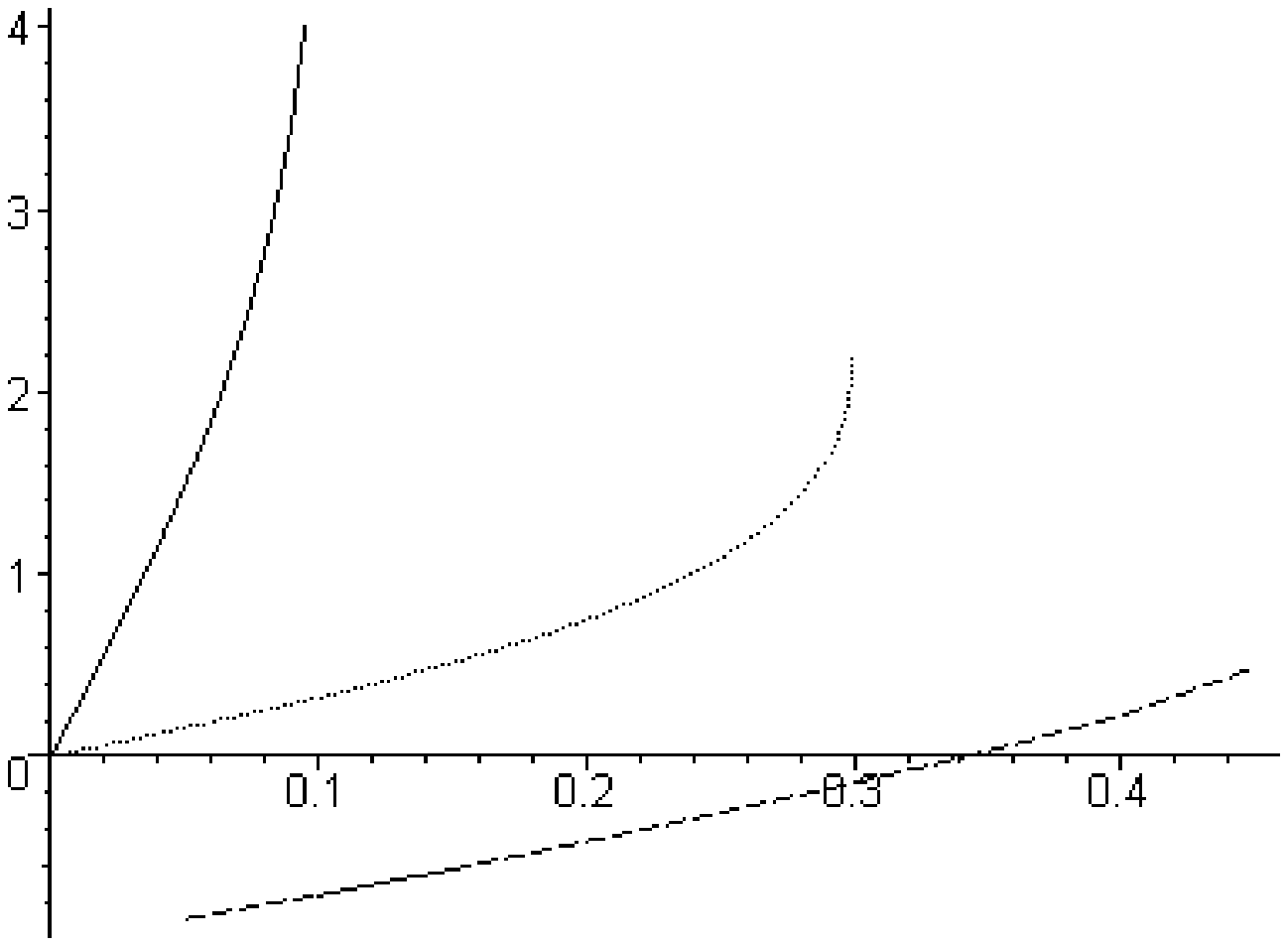,width=0.8\linewidth}
\end{center}
\caption{$R\Omega $ versus $A$ for $M=0.1$ (solid), $0.3$ (dotted), $0.5$
(dashed).}
\label{Figure13}
\end{figure}

Expressing the energy as a function of $R,$ $S$ and $J,$ the angular
velocity can be written as 
\begin{equation}
\Omega =\left( \frac{\partial E}{\partial J}\right) _{S,R}=\left( \frac{%
\partial E}{\partial M}\right) _{S,R}\left( \frac{\partial M}{\partial J}%
\right) _{S,R}=\Omega _{H}\left( \frac{\partial E}{\partial M}\right) _{R,S}
\label{km18}
\end{equation}
where $\Omega _{H}$ is the angular velocity at the event horizon. \ As $%
R\rightarrow \infty $ , $(\partial E/\partial M)_{R,S}\rightarrow 1$ and
thermodynamic chemical potential conjugate to the angular momentum
approaches $\Omega _{H}$ . The $A$ dependence of $\Omega $ is illustrated in
Fig. \ref{Figure13}.

\bigskip

As expected, $\Omega $ decreases with decreasing $A$. \ However the
derivative $(\partial E/\partial M)_{R,S}$ is not positive for all values of 
$A$ once $M$ becomes sufficiently large, or (alternatively) $R$ becomes
sufficiently small. This suggests that some kind of superradiance will take
place for sufficiently small quasilocal boundaries. The numerical
computation required to extrapolate the curve for $M=0.5$ is more intense
than the computational power we have available; however we have explicitly
checked that $\Omega $ vanishes for small $A$ in this case. We shall not
pursue this issue further.

\section{Kerr-AdS Metric}

\bigskip

For the Kerr-AdS metric the action can be written as a linear combination of
\ a volume term, a boundary term and the counterterm (\ref{e7}) in the small-%
$l$ limit: 
\begin{equation}
I=-\frac{1}{16\pi }\{\int_{{\cal M}}d^{4}x\sqrt{-\gamma }\left( {\cal R}%
-2\Lambda \right) +2\int_{\partial {\cal M}}d^{3}x\sqrt{-\gamma }\Theta -%
\frac{4}{l}\int_{\partial {\cal M}}d^{3}x\sqrt{-\gamma }(1+\frac{l^{2}}{4}%
{\cal R}(\gamma ))\}.  \label{ka1}
\end{equation}
\ Under variation of the metric, $\gamma ,$ one obtains: 
\begin{equation}
\delta I=\frac{1}{16\pi }\{[%
\mbox{terms that vanish when the equations of
motion hold}]\delta g_{ij}+[\int_{\partial {\cal M}}\sqrt{-\gamma }%
(P^{ij}+Q^{ij})]\delta \gamma _{ij}\},  \label{ka2}
\end{equation}
where now\ \ 
\begin{equation}
Q^{ij}=\frac{1}{16\pi }(lG^{ij}-\frac{2}{l}\gamma ^{ij}).  \label{ka3}
\end{equation}

Therefore the energy of the system can be written as:\ \ \ \ \ \ \ \ \ \ \ \
\ \ \ 
\begin{equation}
E=E_{1}+E_{2};\ \ \ \ \ \ \ \ \ \ \ \ \ \ \ \ \ \ E_{i}=\int_{B}d^{2}x\sqrt{%
\sigma }\varepsilon _{i},  \label{ka4}
\end{equation}
where $\varepsilon _{1}$ and $\varepsilon _{2}$ are given by\ 

\ \ \ \ \ \ \ 
\begin{eqnarray}
\sqrt{\sigma }\varepsilon _{1} &=&\frac{2}{N}P^{ij}u_{i}u_{j\mbox{ }}
\label{ka5} \\
&=&\frac{Lr\sqrt{1+L^{2}+A^{2}L^{2}+A^{2}-2ML^{2}}}{%
4(L^{2}-A^{2})(L^{2}-A^{2}z^{2})(1+A^{2}z^{2})}  \nonumber \\
&&\times \frac{\lbrack
(L^{2}-A^{2}-ML^{2})A^{2}z^{2}-2A^{2}+A^{2}L^{2}+A^{2}ML^{2}-A^{4}+2L^{2}]}{%
\{(L^{2}+A^{2}L^{2}-A^{2}-2ML^{2}-A^{4})A^{2}z^{2}+A^{2}L^{2}-A^{2}-A^{4}+2A^{2}ML^{2}+L^{2}\}^{1/2}%
}\ \ \ \ \ \ \ \   \nonumber
\end{eqnarray}
\begin{equation}
\begin{tabular}{l}
$\sqrt{\sigma }\varepsilon _{2}=\frac{2}{N}Q^{ij}u_{i}u_{j}$ \\ 
$\ \ \ \ \ \ \ \ =\frac{Lr}{4(L^{2}-A^{2})(1+A^{2}z^{2})^{3}}%
\{(L^{2}+A^{2}L^{2}-A^{4}-A^{2}-2ML^{2})A^{8}z^{8}$ \\ 
$\ \ \ \ \ \ \ \ \ \ +\
(4L^{2}-6ML^{2}+2ML^{2}A^{2}+4L^{2}A^{2}-4A^{2}-4A^{4})A^{6}z^{6}\ \ \ \ \ \
\ \ $ \\ 
$\ \ \ \ \ \ \ \ \ \
+(L^{4}+6ML^{2}A^{4}+A^{2}L^{4}+6A^{2}L^{2}-7A^{2}-A^{4}L^{2}+22ML^{2}A^{2}-4M^{2}L^{4} 
$ \\ 
$\ \ \ \ \ \ \ \ \ \ \ \ \ \ \ \ \ \ \ \ \ \ \ \ \ \ \ \ \ \ \ \ \ \ \ \ \
-7A^{4}+2A^{2}L^{4}M+7L^{2}+2ML^{2})A^{4}z^{4}$ \\ 
$\ \ \ \ \ \ \ \ \ \
+(4A^{2}L^{2}-6A^{2}-6A^{4}-4ML^{2}+4A^{2}M^{2}L^{4}-8A^{4}L^{4}M$ \\ 
$\ \ \ \ \ \ \ \ \ \ \
+6L^{2}-4ML^{2}A^{4}+2A^{2}L^{4}+4ML^{2}A^{2}-10ML^{4}+2L^{4}-14MA^{2}L^{4}-2A^{4}L^{2})A^{2}z^{2} 
$ \\ 
$\ \ \ \ \ \ \ \ \ \
+2MA^{4}L^{4}+2L^{2}-2A^{2}-A^{4}L^{2}+L^{4}A^{2}+4ML^{2}A^{2}+A^{2}L^{2}+L^{4}-2A^{4}+4MA^{2}L^{4}\}\ 
$ \\ 
$\ \ \ \ \ \times \
\{(L^{2}-A^{2}z^{2})[(L^{2}+A^{2}L^{2}-A^{2}-2ML^{2}-A^{4})A^{2}z^{2}+A^{2}L^{2}-A^{2}-A^{4}+2A^{2}ML^{2}+L^{2}]\}^{1/2} 
$%
\end{tabular}
\label{ka6a}
\end{equation}
with $A=a/R$, $M=m/R$ and $L=l/R$.

For $A=0,$ we recover the quasilocal energy of a Schwarzchild-AdS black hole 
\cite{BCM}. Following our previous approach, we first consider special cases
in which the integral in Eq. (\ref{ka4}) can be evaluated. For $R=r_{+}$,
where $r_{+}$ is the radius of \ the horizon, we find that the total energy
can be explicitly computed, yielding 
\begin{equation}
\ E(r_{+})=\frac{l^{2}+3a^{2}+4r_{+}^{2}}{2(l^{2}-a^{2})}l  \label{ka6}
\end{equation}
For\ small values of angular momentum, $E$ can be integrated easily. Then
the total energy for small $A$ is:\ \ \ 
\begin{equation}
E=\frac{R}{L}\{\frac{2+L^{2}}{2}+\frac{2}{3}A^{2}(1+M+\frac{1}{L^{2}})-\sqrt{%
1+A^{2}-2M+\frac{1+A^{2}}{L^{2}}}[1+\frac{A^{2}}{6}(1+2M-\frac{4}{L^{2}})%
]+O(A^{4})\}  \label{ka7}
\end{equation}
\bigskip When $A=0$ the energy is given by 
\begin{equation}
E=R\left( \frac{1}{L}+\frac{L}{2}-\sqrt{1+\frac{1}{L^{2}}-2M}\right)
\label{ka7a}
\end{equation}
which differs from the expression obtained using background subtraction
methods \cite{BCM}. \ 

For general values of the parameters, however, we cannot analytically
integrate (\ref{ka4}) exactly, and so we resort to numerical integration. \
To do this entails several restrictions on the parameters. First, the
Kerr-AdS solution is defined only for $L>A$. Furthermore, since the
quasilocal surface must be outside of the horizon, $r_{+}/r$ must be real
and less than unity. Note that there is no longer a restriction on the sign
of the induced Ricci scalar. These criteria yield this fact that if one
choose two of the three parameters $M$ , $L$ and $A$, then the allowed
values of the third parameter should obeys the following equations:


\begin{eqnarray}
&&\mbox{for given values of }A\mbox{ and }L,\mbox{ }M_{\mbox{crit}}\leq
M\leq \frac{(1+A^{2})(1+L^{2})}{2L^{2}},\mbox{ if }2M>(1+A^{2}),  \nonumber
\\
&&\mbox{for given values of }A\mbox{ and }M,\mbox{ \ }L_{\mbox{crit}}\leq
L\leq \sqrt{\frac{(1+A^{2})}{2M-(1+A^{2})}}\mbox{, if }2M>(1+A^{2})\mbox{ },
\label{kadscond} \\
&&\mbox{for given values of }M\mbox{ and }L,\mbox{ }\sqrt{\frac{2ML^{2}}{%
1+L^{2}}-1}\leq A\leq A_{\mbox{crit}}\mbox{ if }2M>(1+L^{2}),  \nonumber
\end{eqnarray}


\noindent where $L_{\mbox{crit}}=\frac{l_{\mbox{crit}}}{R},$ $A_{\mbox{crit}%
}=\frac{a_{\mbox{crit}}}{R},$ and $M_{\mbox{crit}}=\frac{m_{\mbox{crit}}}{R}$
given by (\ref{kadscrit}). For $M\neq 0.5$ and $2M<1+A^{2}$ there is no
upper bound for $L,$ and for $2M<(1+L^{2}),$ the lower bound for $A$ is zero$%
.$

\bigskip

\begin{figure}[tbp]
\begin{center}
\epsfig{file=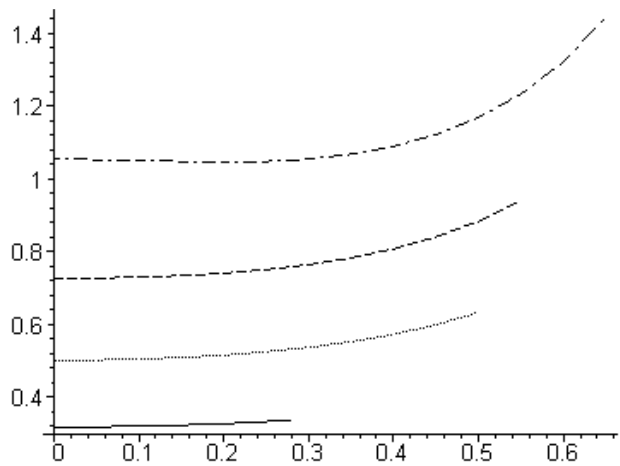,width=0.8\linewidth}
\end{center}
\caption{$E/R$ versus $A$ for $L=1,$ $M=0.3$ (solid), $0.5$ (dotted), $0.7$
(dashed), $0.9$ (dot-dashed).}
\label{Figure 14}
\end{figure}

\begin{figure}[tbp]
\begin{center}
\epsfig{file=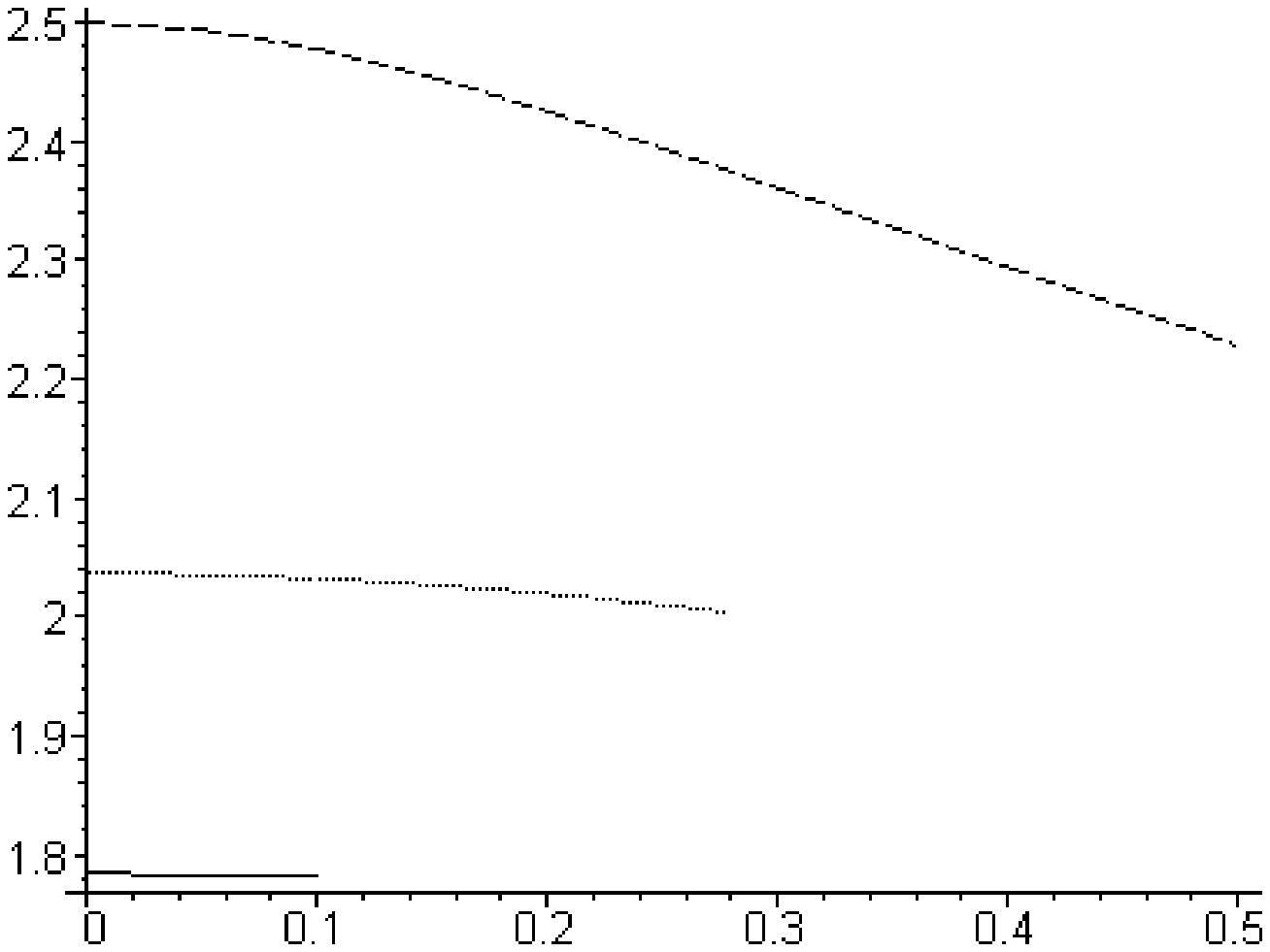,width=0.8\linewidth}
\end{center}
\caption{$E/R$ versus $A$ for $L=5,$ $M=0.1$ (solid) $0.3$ (dotted), $0.5$
(dashed). }
\label{Figure15}
\end{figure}

In the next four figures we plot the quasilocal energy as a function of $A$
and $M$. The $A$-dependence of the energy is given in Figs. \ref{Figure 14}
and \ref{Figure15}. We see that for small values of \ $L$ ($L\sim 1$) the
energy is a slowly increasing function of $A.$ For larger values of $\ L,$
as in the case of Kerr metric, for $M<0.5$ the energy decreases as $A$
increases.

\bigskip

\bigskip

\begin{figure}[tbp]
\begin{center}
\epsfig{file=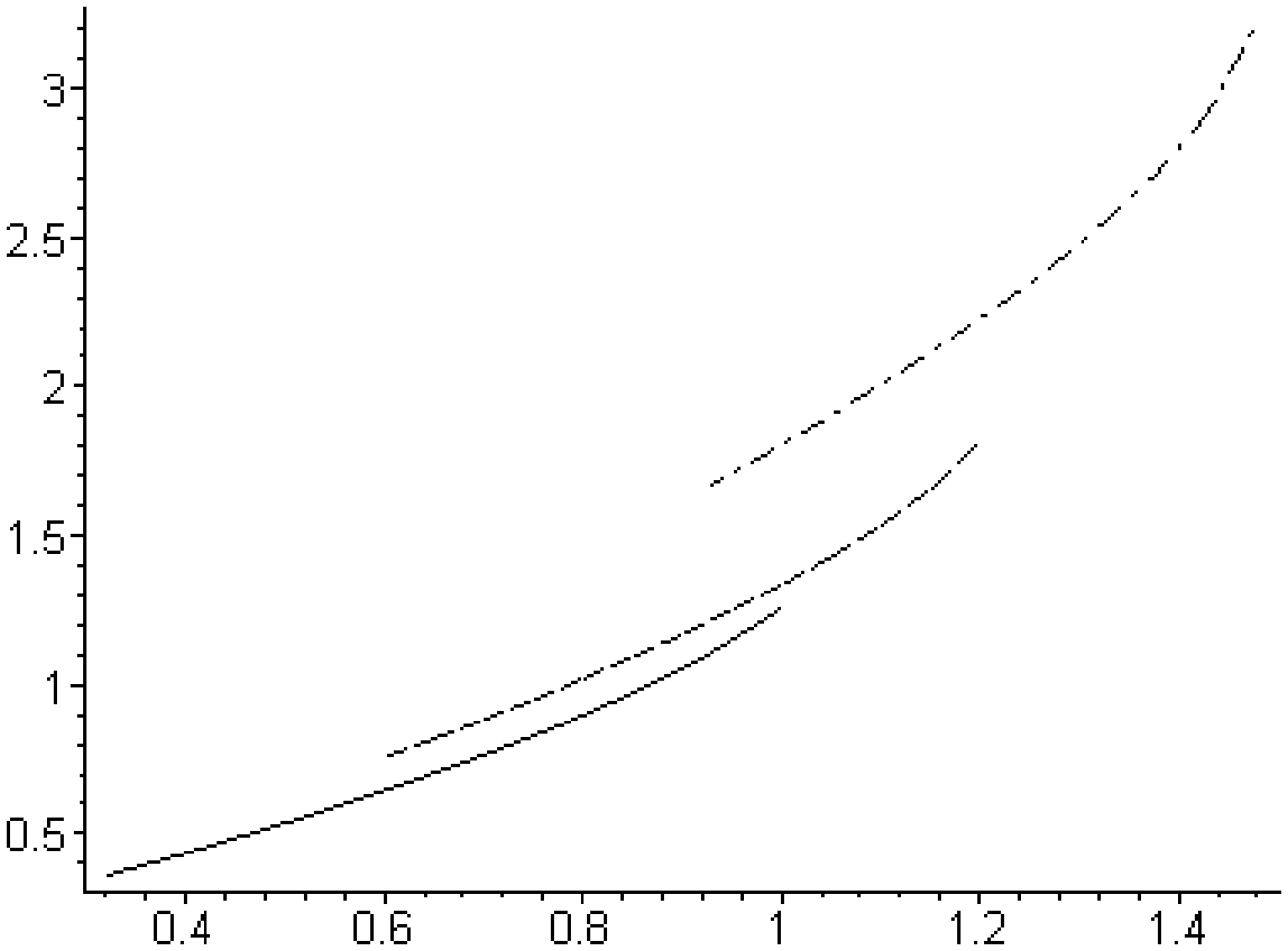,width=0.8\linewidth}
\end{center}
\caption{$E/R$ versus $M$ \ for $L=1$, $A=0.3$ (solid), $0.5$ (dashed), $0.7$
(dot-dashed)$.$}
\label{Figure 16}
\end{figure}

\begin{figure}[tbp]
\begin{center}
\epsfig{file=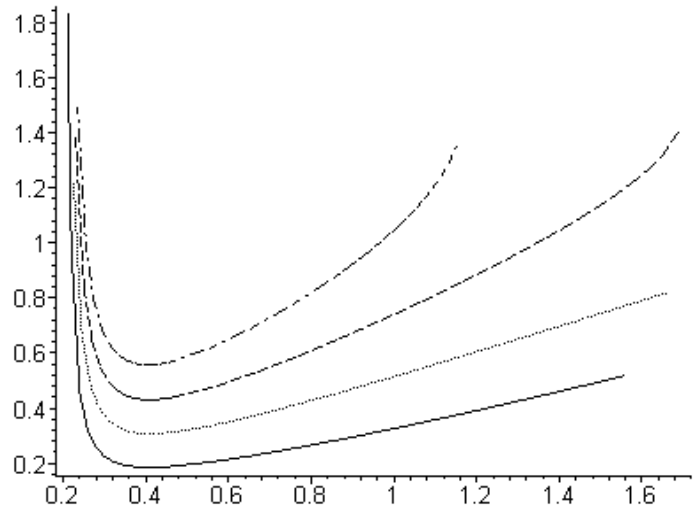,width=0.8\linewidth}
\end{center}
\caption{$E/R$ versus $L$ for $A=0.2$, $M=0.3$ (solid), $0.5$ (dotted), $.7$
(dashed), $0.9$ (dot-dashed).}
\label{Figure 17}
\end{figure}

Figures \ref{Figure 16} and \ref{Figure 17} show the $M$-dependence and $L$%
-dependence of the energy. For large values of $L$ the energy is asymptotic
to a linear function of $L$ $.$ This is due to the fact that the counterterm
used for Kerr-AdS is proportional to $L$ for large values of $L$ .

\bigskip

\bigskip

As in the case of Kerr metric we consider the quasilocal conserved charges
of the Kerr-AdS metric which are the mass and $\varphi $ component of
angular momentum of the system defined in eqns. (\ref{e19},\ref{e20}). \ The 
$\varphi $ component of angular momentum due to the counterterm is zero and
one can evaluate the total angular momentum as

\begin{equation}
J_{\varphi }=am(1-\frac{a^{2}}{l^{2}})^{-2},  \label{ka8}
\end{equation}
which is therefore valid for any quasilocal surface of fixed $r=R$, and
yields the angular momentum of the Kerr metric as $l$ goes to infinity.

\begin{figure}[tbp]
\begin{center}
\epsfig{file=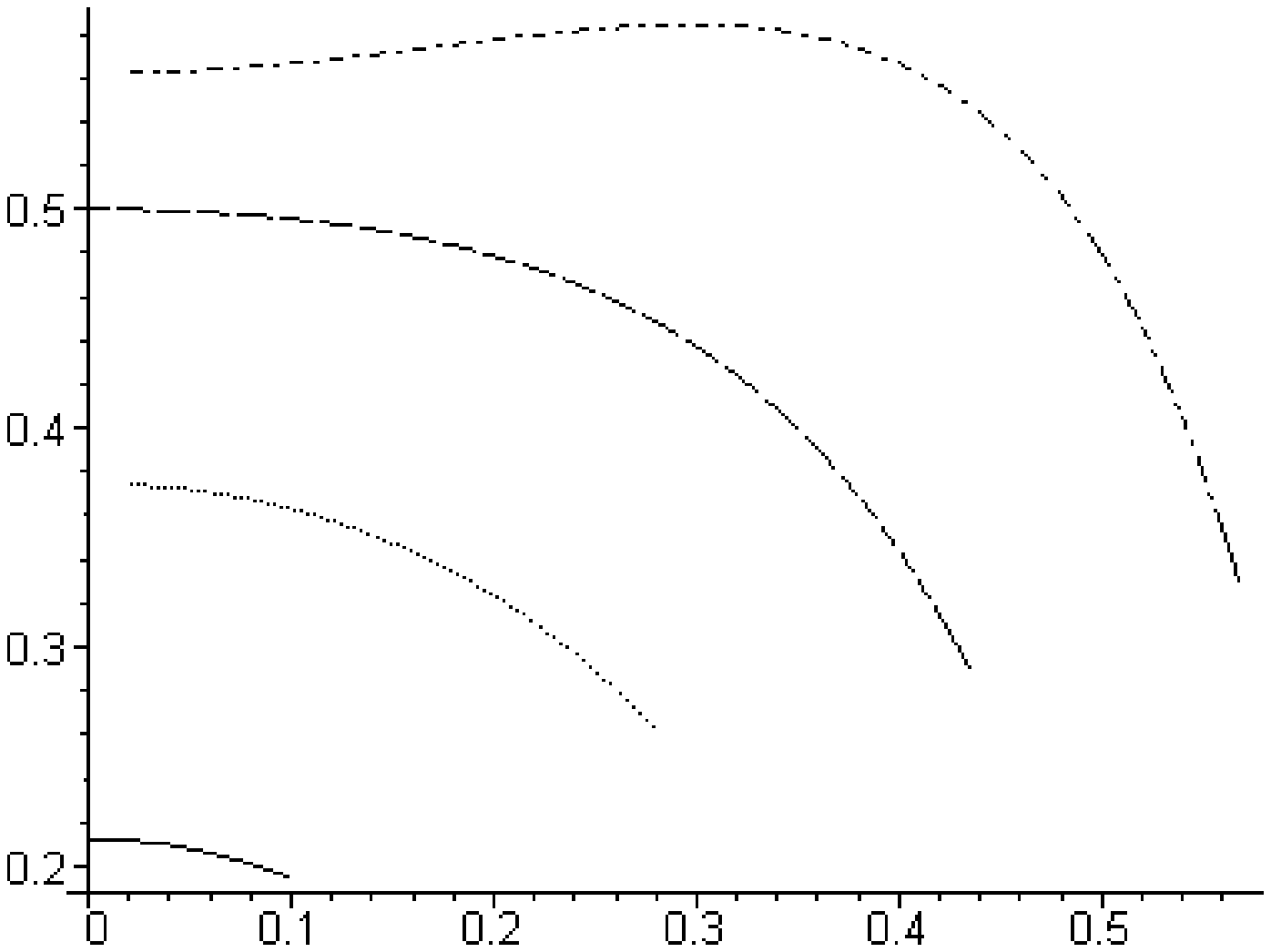,width=0.8\linewidth}
\end{center}
\caption{${\cal M}/R$ versus $A$ for $L=1,$ $M=0.1$ (solid), $0.3$ (dotted), 
$0.5$\ (dashed), $0.7$ \ (dot-dashed).}
\label{Figure 18}
\end{figure}

\begin{figure}[tbp]
\begin{center}
\epsfig{file=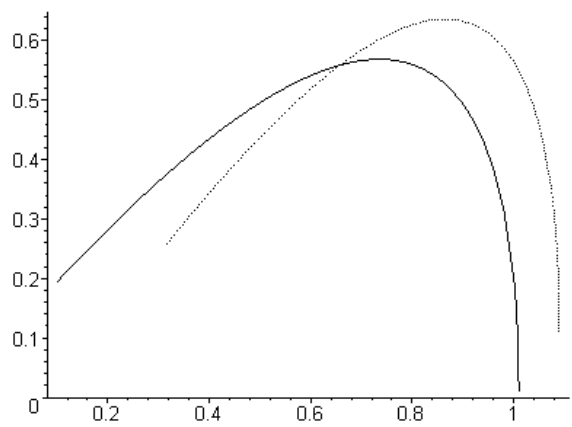,width=0.8\linewidth}
\end{center}
\caption{${\cal M}/R$ versus $M$ for $L=1,$ $A=0.3$ (solid), $0.5$\
(dotted). }
\label{Figure 19}
\end{figure}

The total mass of the system, ${\cal M}$, is somewhat more complicated to
evaluate. For $R=r_{+}$ \ it is

\begin{equation}
{\cal M}\left( r_{+}\right) {\cal =}\frac{a^{2}}{2r_{+}}(1+\frac{r_{+}^{2}}{%
l^{2}})(1-\frac{a^{2}}{l^{2}})^{-2},  \label{ka9}
\end{equation}
\bigskip where we have used the normalized Killing vector $\zeta
^{a}=(\partial /\partial t)^{a}/$ $\Xi $ instead of $(\partial /\partial
t)^{a}$ in Eq. (\ref{e19}) so that the conserved quantities at infinity
generate the $SO(3,2)$ algebra, in agreement with the conventions of Ref. 
\cite{kostperry}. The mass ${\cal M}$ as $r$ goes to\ infinity then becomes 
\begin{equation}
{\cal M}_{\infty }=\frac{m}{\Xi ^{2}},  \label{ka9a}
\end{equation}
which differs from that of eq.(\ref{kadsmj}) by this normalization factor.
The $A$ and $M$-dependence of ${\cal M}$ are illustrated in Figs. \ref
{Figure 18} and \ref{Figure 19}.

\bigskip

\bigskip

\bigskip

Adopting the same stance as for the Kerr black hole, we express the energy
as a function of $R,$ $J$ and $S$. Noting that the entropy is $S=\frac{\pi }{%
\Xi }(r_{+}^{2}+a^{2})$, the temperature can be written as 
\begin{equation}
T=\left( \frac{\partial E}{\partial S}\right) _{J,R}=\left( \frac{\partial E%
}{\partial M_{\infty }}\right) _{J,R}\left( \frac{\partial S}{\partial
M_{\infty }}\right) _{J,R}^{-1}=\frac{\kappa _{+}}{2\pi R}\left( \frac{%
\partial E}{\partial M_{\infty }}\right) _{R,J}  \label{ka10}
\end{equation}
where $\kappa _{+}$ is the surface gravity at the event horizon 
\begin{equation}
\kappa _{+}=\frac{3r_{+}^{4}+r_{+}^{2}(a^{2}+l^{2})-a^{2}l^{2}}{%
2l^{2}r_{+}(r_{+}^{2}+a^{2})}.  \label{ka11}
\end{equation}
and $M_{\infty }={\cal M}_{\infty }/R$.

\begin{figure}[tbp]
\begin{center}
\epsfig{file=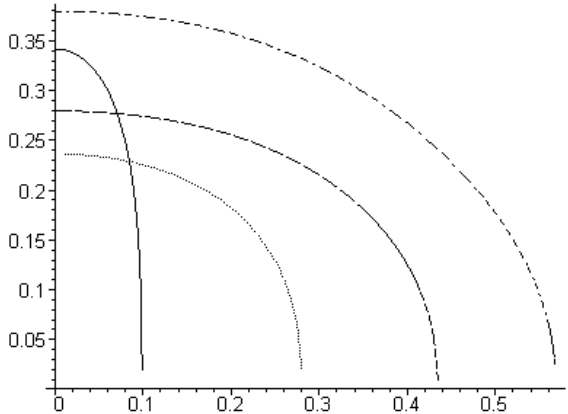,width=0.8\linewidth}
\end{center}
\caption{$\ RT$ versus $A$ for $L=1,$ $M=0.1$ (solid), $0.3$ (dotted), $0.5$%
\ (dashed), $0.7$ \ (dot-dashed).}
\label{Figure 20}
\end{figure}

\begin{figure}[tbp]
\begin{center}
\epsfig{file=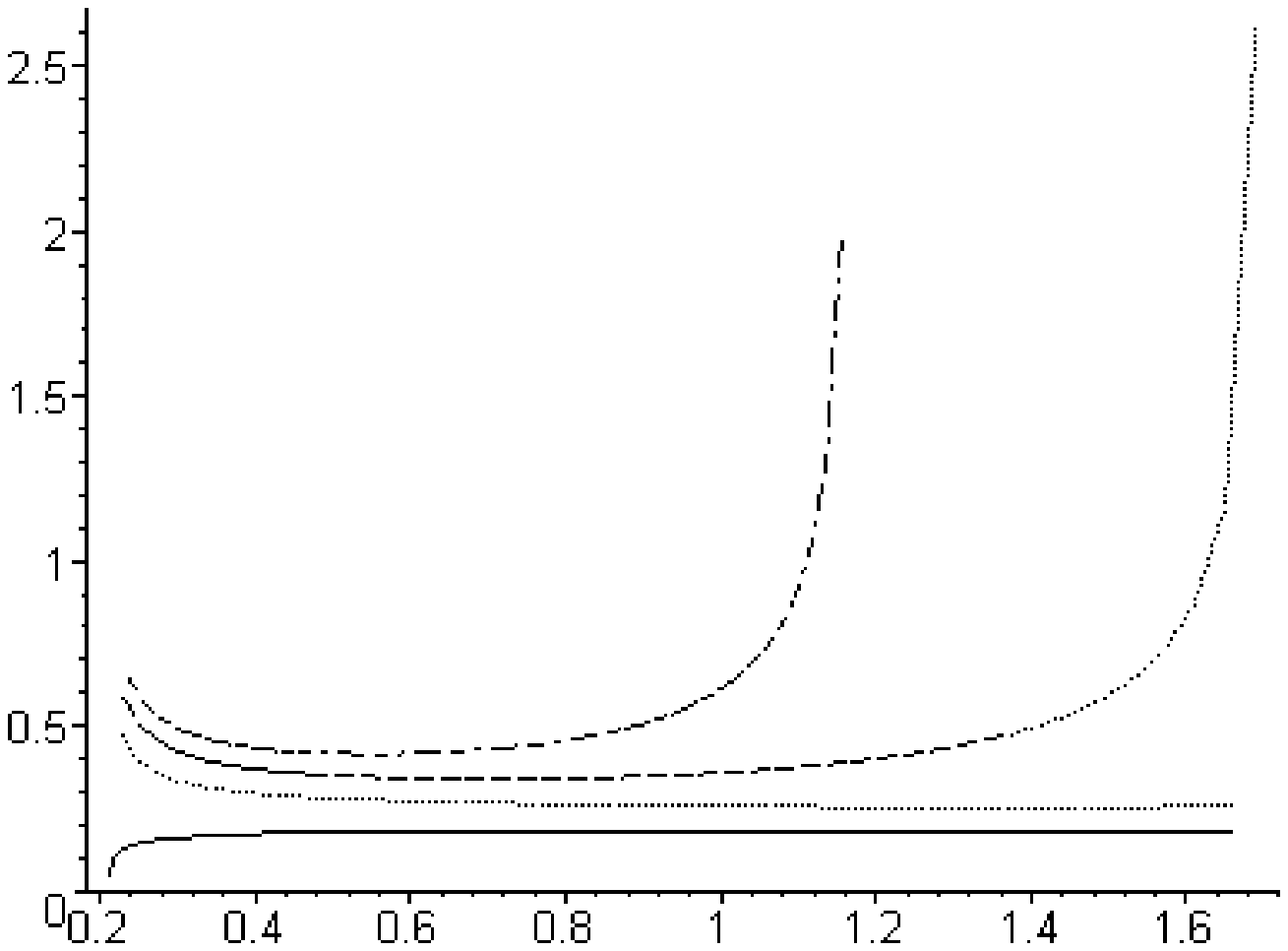,width=0.8\linewidth}
\end{center}
\caption{$RT$ versus $L$ for $A=0.2,$ $M=0.3$ (solid), $0.5$ (dotted), $0.7$%
\ (dashed), $0.9$ \ (dot-dashed).}
\label{Figure 21}
\end{figure}

\begin{figure}[tbp]
\begin{center}
\epsfig{file=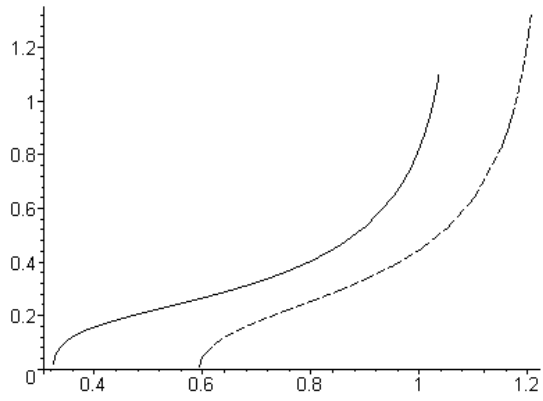,width=0.8\linewidth}
\end{center}
\caption{$RT$ versus $M$ for $L=1,$ $A=0.3$ (solid), $0.5$ (dashed).}
\label{Figure 22}
\end{figure}

As for the Kerr case, the temperature must be computed numerically. We find
(fig. \ref{Figure 20}) that for small values of $M$ the temperature is a
weakly decreasing function of $A$ , and (similar to the Kerr case) it
decreases rapidly as $A$ approaches its maximum allowed value. As $M$
increases, the $A$-dependence of $T$ \ becomes more relevant. The $L$%
-dependence of the temperature can be seen from fig. \ref{Figure 21}. Note
that the value of $T$ due to the counterterm is very small compared to its
value due to the boundary term. In contradistinction to the energy, we find
that $T$ is an initially increasing function of $L$ for small $M$. For l$%
\arg $e values of $L$ \ ($L>100$) the change in $T$ due to the boundary term
is negligible and $T$ increases as $L$ increases. The $M$-dependence of $T$
\ is plotted in \ref{Figure 22}.

\bigskip

\bigskip \bigskip

\bigskip

\bigskip

\begin{figure}[tbp]
\begin{center}
\epsfig{file=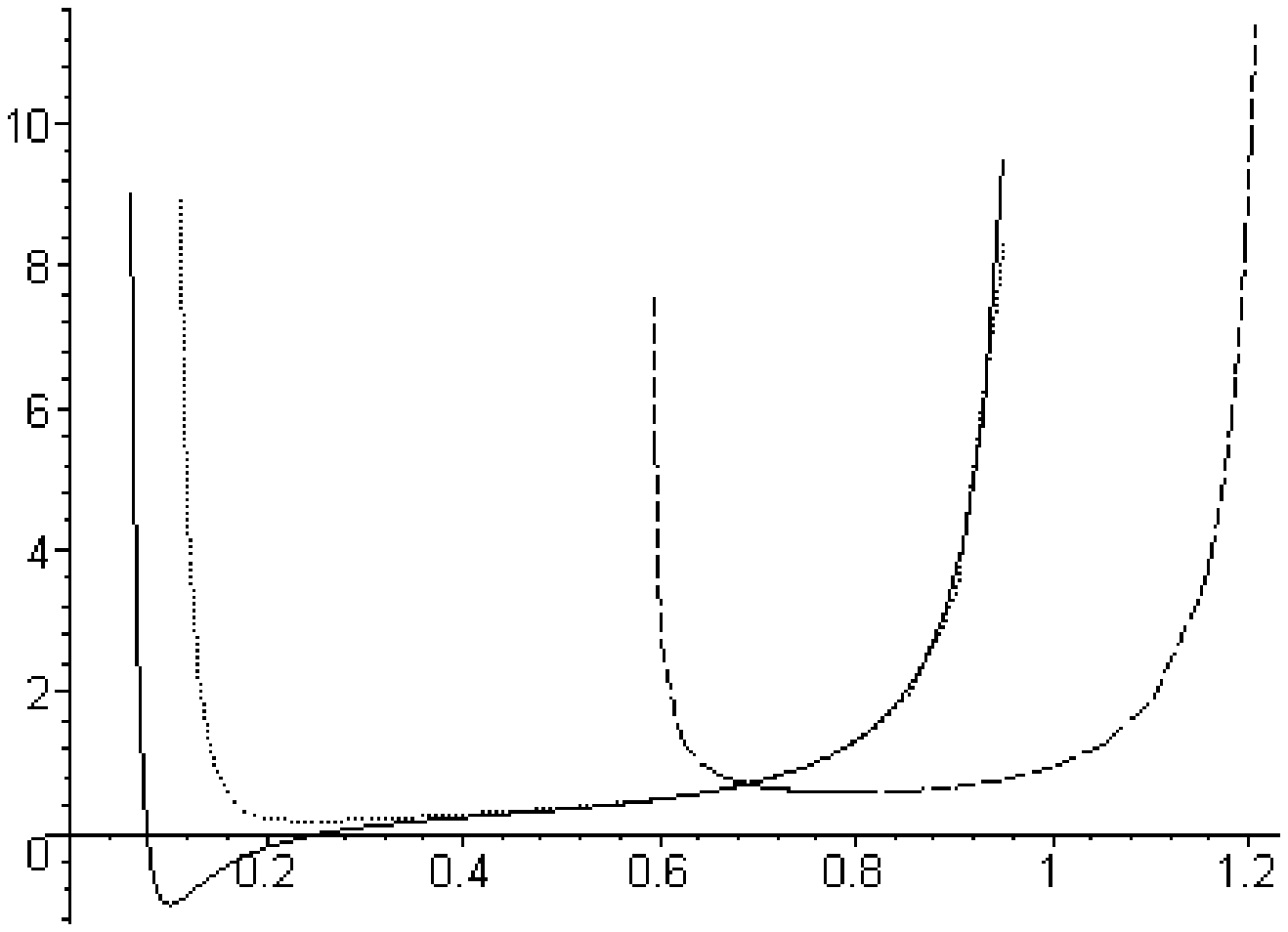,width=0.8\linewidth}
\end{center}
\caption{$R(\partial T/\partial M_{\infty })_{J,R}$ versus $M$ for $L=1,$ $%
A=0.05$ (solid), $0.1$ (dashed), $0.5$ (dashed).}
\label{Fig23}
\end{figure}

In figure \ref{Fig23} we plot the $M-$dependence of $(\frac{\partial T}{%
\partial M})_{J,R}.$ Again this figure is in consistent with that of Kerr
case, and shows that for $A=0.05$, $T(M,J,R)$ has a maximum at $A\simeq
0.078 $ and a minimum at $A\simeq 0.25.$ We find for small values of $A$ a
single stable phase consisting of a small black hole. As the temperature
increases, a second stable phase emerges consisting of a large black hole,
along with an \ intermediate-mass unstable phase. \ This latter phase
coalesces with the small black hole at an even larger temperature, beyond
which there exists only the single stable large black hole phase. \ As $A$
increases the small and unstable phases are present for continually smaller
temperature ranges. They eventually vanish for $A$ sufficiently large,
leaving only the single large black hole phase. The heat capacity at
constant surface area $4\pi R^2$ can be evaluated numerically, and the
results, plotted as functions of $A$ and $M$, are shown in Figs. \ref{Figure
24} and \ref{Figure 25}.

\bigskip

\begin{figure}[tbp]
\begin{center}
\epsfig{file=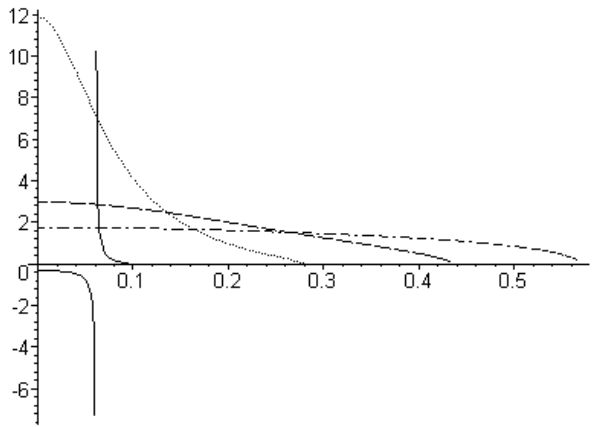,width=0.8\linewidth}
\end{center}
\caption{$R^{-2}C_{R}$ versus $A$ for $L=1,$ $M=0.1$ (solid), $0.3$
(dotted), $0.5$\ (dashed), $0.7$ \ (dot-dashed).}
\label{Figure 24}
\end{figure}

\bigskip

\begin{figure}[tbp]
\begin{center}
\epsfig{file=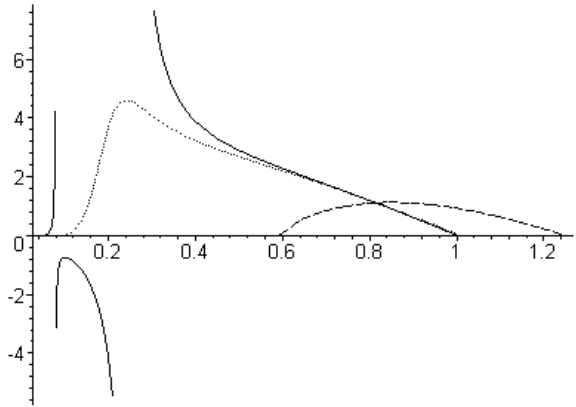,width=0.8\linewidth}
\end{center}
\caption{$R^{-2}C_{R}$ versus $M$ for $L=1,$ $A=0.05$ (solid), $0.1$
(dotted), $0.5$\ (dashed).}
\label{Figure 25}
\end{figure}

The stability analysis for these black holes is in qualitative agreement
with that of Caldarelli et. al. \cite{Cald}, who recently performed a
thorough analysis of the thermodynamic properties of Kerr-Newman AdS black
holes in the $R\rightarrow \infty $ limit. \ However they computed a
generalization of the Smarr formula for the AdS case in terms of ${\cal M}%
_{\infty }$, whereas we are considering the quasilocal energy $E$ in (\ref
{ka4}) as a function of $R$, $S$, and $J$. \ As a consequence the
temperature they derive does not redshift to zero in the large-$R$ limit,
unlike the expression in (\ref{ka10}) which does, although the explicit form
differs from that of the Tolman factor due to the quasilocal surface not
being an isotherm when $J\neq 0$.

\begin{figure}[tbp]
\begin{center}
\epsfig{file=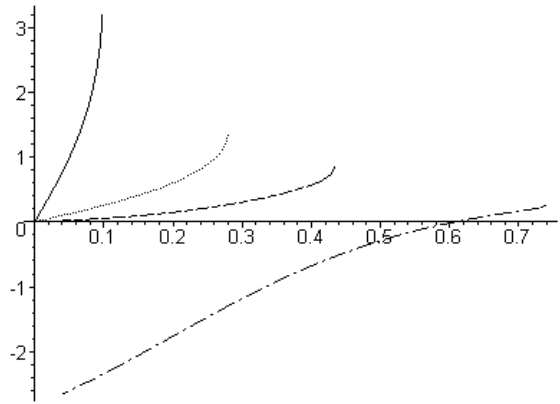,width=0.8\linewidth}
\end{center}
\caption{$R\Omega $ versus $A$ for $L=1,$ $M=0.1$ (solid), $.3$ (dotted), $%
.5 $\ (dashed), $1.0$ (dot-dashed).}
\label{Figure 26}
\end{figure}

The angular velocity can be written as:

\begin{equation}
\Omega =\left( \frac{\partial E}{\partial J}\right) _{S,R}=\left( \frac{%
\partial E}{\partial M_{\infty }}\right) _{S,R}\left( \frac{\partial
M_{\infty }}{\partial J}\right) _{S,R}=\Omega _{E}\frac{1}{R}\left( \frac{%
\partial E}{\partial M_{\infty }}\right) _{S,R}  \label{ka12}
\end{equation}
where $\Omega _{E}$ is the angular velocity of the Einstein universe: 
\begin{equation}
\Omega _{E}=\frac{a(1+\frac{r_{+}^{2}}{l^{2}})}{r_{+}^{2}+a^{2}}=\Omega
_{H}-\Omega _{\infty }  \label{ka13}
\end{equation}
with $\Omega _{H}=a(1-a^{2}/L^{2})/(r_{+}^{2}+a^{2})$ the angular velocity
of the hole at event horizon, and \ $\Omega _{\infty }=-a/l^{2}$ is the
angular velocity at infinity. As $R\rightarrow \infty $, $(\partial
E/\partial M_{\infty })_{S,R}\rightarrow 1$, and the results of Ref. \cite
{Cald}\ are recovered. The $A$ dependence of the angular velocity can be
seen from \ref{Figure 26}. As in the Kerr case, we find that $(\partial
E/\partial M_{\infty })_{S,R}$ is not positive once $M$ is sufficiently
large. Again, although the numerical computation required to extrapolate the
curve for $M=1$ is more intense than the computational power we have
available, we have checked explicitly that $\Omega $ vanishes for small $A$
in this case.

\bigskip

\section{2+1 BTZ Black Hole}

We can analyze the (2+1)-dimensional axially symmetric black hole solution
of Einstein equation with a negative cosmological constant $\Lambda
=-1/l^{2} $ obtained by Banados et.al. using methods similar to those given
above. Written in stationary coordinates the BTZ metric is \cite{BTZ} 
\begin{equation}
ds^{2}=-N^{2}(r)dt^{2}+f^{-2}(r)dr^{2}+r^{2}[V^{\phi }(r)dt+d\phi ]^{2},
\label{btz1}
\end{equation}
with 
\begin{equation}
N^{2}(r)=f^{2}(r)=-m+\left( \frac{r}{l}\right) ^{2}+\left( \frac{j}{2r}%
\right) ^{2},\mbox{ \ \ }V^{\phi }(r)=-\frac{j}{2r^{2}}.\mbox{\ }.
\label{btz2}
\end{equation}

The action becomes 
\begin{equation}
I=-\frac{1}{8\pi }\{\int_{{\cal M}}d^{4}x\sqrt{-\gamma }\left( {\cal R}%
-2\Lambda \right) +2\int_{\partial {\cal M}}d^{3}x\sqrt{-\gamma }\Theta -%
\frac{2}{l}\int_{\partial {\cal M}}d^{3}x\sqrt{-\gamma }\},  \label{btz3}
\end{equation}
where we drop the counterterm corresponding to the induced Ricci scalar
since it makes no contribution at infinity in (2+1) dimensions. The analogue
of eq. (\ref{ka3}) is 
\begin{equation}
Q^{ij}=\frac{\sqrt{-\gamma }}{2\pi }(-\frac{1}{l}\gamma ^{ij}).  \label{btz4}
\end{equation}
from which it is easy to show that the energy is

\begin{equation}
E=2\left( \frac{r}{l}-\sqrt{-m+\frac{r^{2}}{l^{2}}+\frac{j^{2}}{r^{2}}}%
\right) .
\end{equation}

This expression is identical to the one given in Ref. \cite{BCM}, in which
the reference spacetime was taken to be that of a zero-mass black hole. \ We
see here that the counterterm prescription at arbitrary $r$ is equivalent to
this choice of spacetime background, and so all of the results of Ref. \cite
{BCM} will hold.

\bigskip

\section{Closing Remarks}

\bigskip

Our investigations of the boundary-induced counterterm prescription (\ref{e7}%
) at finite distances has allowed us to obtain a number of interesting
results, which we recapitulate here.

We have been able to compute for the first time the energy, conserved mass,
and angular momentum quasilocally for arbitrary values of the parameters of
the Kerr and Kerr-AdS solutions, apart from the mild reality restrictions 
(\ref{kerrconds}) and
(\ref{kadscond}). \ These quasilocal quantities are intrinsically calculable
at fixed $r=R$ without any reference to a background spacetime, a marked
improvement over the methods given in Ref. \cite{Martinez}, which were
necessarily restricted to small values of the rotation parameter. We
recovered these results, and were further able to show that the conjecture
that $E\left( r_{+}\right) =2M_{i}$ (where $M_{i}$ is the irreducible mass
of the Kerr black hole) was not correct, but instead was valid only for
small $A$. \ We also found that quasilocal angular momentum equals its value
at infinity at any fixed value $r=R$. \ The remarkable result holds because
the counterterms integrate to zero, and not because they identically vanish.

We found that the entropy $S$ and angular momentum $J$ were not given by
surface integrals over quasilocal boundary data, but rather were constants
dependent on the parameters $a,m$ (and $l$) of the black hole. We therefore
chose to regard the interior of the quasilocal surface as a thermodynamic
system whose energy $E$ was a function of $S$ and $J$. A stability analysis
yielded results that were in qualitative agreement with previous
investigations carried out for either the non-rotating case \cite{BCM}, or
at infinity \cite{Davies,Cald}. This stability analysis is, of course,
local, and phase transitions to other spacetimes of lower free energy might
exist. For the BTZ black hole the boundary countermterm turned out to be
formally identical to a contribution from a reference spacetime of zero mass 
\cite{BCM}, and so the thermodynamics was identical in the two cases. It
would be interesting to repeat these investigations for a different choice
of quasilocal boundary, such as an isotherm.

The counterterm (\ref{e7}) reduces to the two cases given in eqs. (\ref{km1}%
) and (\ref{ka1}) the large-$l$ and ${\cal O}\left( l\right) $ limits
respectively. \ These two regimes are quite distinct except when the
boundary curvature scalar ${\cal R}(\gamma )\sim 1/l$ , at which point one
might expect that other boundary curvature invariants besides the minimal
number we have considered here would be relevant. However we have not found
it necessary to include such terms.

Indeed, boundary counterterms evidently have a robust degree of
applicability, going well beyond that originally proposed from their origins
in the AdS/CFT correspondence conjecture. \ In some ways this is not too
surprising, since one need have no {\it {a-priori}} knowledge of the
conjecture to introduce such terms. \ However we have found the somewhat
unexpected result that inclusion of the minimal number of intrinsic boundary
terms needed to make quantities well defined at infinity is sufficient to
compute the thermodynamics of both asymptotically flat and asymptotically
AdS black holes for virtually all possible values of the rotation parameter
and cosmological constant that leave the quasilocal boundary well defined. \
This goes well beyond what is possible with background subtraction methods,
and perhaps suggests that a deeper physical meaning be ascribed to such
terms.

{\it \bigskip }

{\it {\Large Acknowledgments} }

M.H.D. is grateful for the hospitality of the Department. of Physics of
Waterloo University, where this work was performed, and for partial support
from the Research council of Shiraz University of Iran. We would like to
thank I. Booth for interesting correspondence. This work was supported by
the Natural Sciences and Engineering Research Council of Canada.{\it \ }

\end{document}